\documentclass{article}%
\usepackage{amsmath}
\usepackage{amsfonts}
\usepackage{amssymb}
\usepackage{graphicx}%
\newcommand{\bpartial}{\mathop{\partial\kern -4pt\raisebox{.8pt}{$|$}}}
\newcommand{\bra}{\mathopen{[\kern-1.6pt[}}
\newcommand{\ket}{\mathclose{]\kern-1.5pt]}}
\newcommand{\bbra}{\mathopen{[\kern-2.2pt[\kern-2.3pt[}}
\newcommand{\bket}{\mathclose{]\kern-2.1pt]\kern-2.3pt]}}

\makeindex
\begin{document}

\title {\large{ \bf CLASSICAL $r$-MATRICES OF TWO AND THREE DIMENSIONAL LIE
SUPER-BIALGEBRAS AND THEIR POISSON-LIE SUPERGROUPS}}

\vspace{3mm}

\author {  \small{ \bf A. Eghbali }\hspace{-2mm}{ \footnote{ e-mail: a.eghbali@azaruniv.edu}} { \small
and}
\small{ \bf A. Rezaei-Aghdam }\hspace{-2mm}{ \footnote{Corresponding author. e-mail:
rezaei-a@azaruniv.edu}} \\
{\small{\em Department of Physics, Faculty of Science, Azarbaijan Shahid Madani University, }}\\
{\small{\em  53714-161, Tabriz, Iran  }}}

\maketitle

\begin{abstract}
We obtain the classical r-matrices of two and three dimensional
Lie super-bialgebras. We thus classify all two  and three
dimensional coboundary Lie super-bialgebras and their types
(triangular, quasi-triangular, or factorable). Using the Sklyanin
superbracket, we then obtain the super Poisson structures on the
related Poisson-Lie supergroups.
\end{abstract}

{\bf keywords:} {Lie super-bialgebra; Poisson-Lie supergroup; Classical r-matrix.}

\newpage

\section {\large {\bf Introduction}}

\smallskip

From  mathematical point of view, Lie bialgebras were first
introduced by Drinfel'd  as algebraic structures and classical
limit of underlying quantized enveloping algebras ({\it quantum
groups}) \cite{Drin}. In particular, every deformation of a
universal enveloping algebra induces a Lie bialgebra structure on
the underlying Lie algebra. Conversely, as it has been shown in
Ref. \cite{Etin} that each Lie bialgebra admits quantization. So
the classification of Lie bialgebras can be seen as the first
step in the classification of quantum groups. On the other hand,
from the physical point of view, the theory of classical
integrable systems naturally relates to the geometry and
representation theory of Poisson-Lie groups and the corresponding
Lie bialgebras and their classical r-matrices (see, for example,
Ref. \cite{Kosmann}). Furthermore, Lie bialgebras and Poisson-Lie
groups have applications in the theory of Poisson-Lie $T$-dual
sigma models \cite{K.S1} and $N=2$ superconformal field theories
\cite{Parkh}. In the same way, Lie super-bialgebras \cite{N.A},
as the underlying symmetry algebras, play an important role in the
integrable structure of $AdS/CFT$ correspondence \cite{Bs}, as
well as in Toda models on Lie superalgebras \cite{EVANS}.
Similarly, one can consider Poisson-Lie $T$-dual sigma models on
Poisson-Lie supergroups \cite{ER}. In this way and by considering
that there is a universal quantization for Lie super-bialgebras
\cite{Geer}, the classification of Lie super-bialgebras
(especially low dimensional Lie super-bialgebras) is important
role from both physical and mathematical point of view. In Refs.
\cite{ER1} and \cite{ER4} we classify all two and three
dimensional Lie super-bialgrbras. In the present paper, following
\cite{ER1} and \cite{ER4},  we find the classical r-matrices of
these Lie super-bialgebras and determine their types (triangular,
quasi-triangular or factorizable). Furthermore, we obtain super
Poisson structures on the related   Poisson-Lie supergroups.

\smallskip
\goodbreak

The paper is organized as follows. In section two, we give some
basic definitions and notations on coboundary  Lie
super-bialgebras and Manin super triple. In section three, we
list the decomposable and indecomposable  Lie superalgebras and
their related Lie super-bialgebras \cite{{ER1}, {ER4}}.  In
section four, using the adjoint and the matrix representation we
calculate classical r-matrices and determine two and three
dimensional coboundary (triangular, quasi-triangular or
factorizable) Lie super-bialgebras. In section five, we first
calculate invariant supervector fields and then using the Sklyanin
superbracket we obtain super Poisson structures  on the related
two and three dimensional  Poisson-Lie supergroups.

\vspace{8mm}

\section {\large {\bf Definitions and notations }}

In the present paper, we apply DeWitt notations for supervector
spaces, supermatrices, etc \cite{D}. Let us  recall some basic
definitions and propositions on Lie super-bialgebras
\cite{{N.A},{ER1},{RHR}}.

\smallskip
\goodbreak

{\it Definition}: A {\em Lie superalgebra} ${\bf g} $ is a graded
vector  space ${\bf g}={\bf g}_B \oplus {\bf g}_F$ with gradings
$grade({\bf g}_B)=0$ and $grade({\bf g}_F)=1$  so that Lie
bracket satisfies the super antisymmetric and super Jacobi
identities, i.e., in the graded basis $\{X_i\}$ of ${\bf g}$ we
have\footnote{Note that the bracket of one boson with one boson
or one fermion is usual commutator, but for one fermion with one
fermion is anticommutator. Furthermore, we identify grading of
indices by the same indices in the power of (-1), for example
$grading(i)\equiv i$; this is the notation that DeWitt applied in
 \cite{D} .}
\begin{equation}
[X_i , X_j] = {f^k}_{ij} X_k,
\end{equation}
and
\begin{equation}
(-1)^{i(j+k)}{f^m}_{jl}{f^l}_{ki} + {f^m}_{il}{f^l}_{jk} +
(-1)^{k(i+j)}{f^m}_{kl}{f^l}_{ij}=0,
\end{equation}
and hence
\begin{equation}
{f^k}_{ij}=-(-1)^{ij}{f^k}_{ji}.
\end{equation}
Furthermore, we have
\begin{equation}
{f^k}_{ij}=0, \hspace{10mm} if \hspace{5mm} grade(i) +
grade(j)\neq grade(k)\hspace{3mm} (mod 2).
\end{equation}
Let ${\bf g}$ be a finite-dimensional Lie superalgebra
and ${\bf g}^\ast$ be its dual superspace with respect to a
non-degenerate canonical pairing $( . ~,~. )$ on ${\bf g}^\ast
\oplus {\bf g}$.

\smallskip

{\it Definition}: A {\em Lie super-bialgebra} structure on a Lie
superalgebra ${\bf g}$ is a  linear map $\delta : {\bf g
}\longrightarrow {\bf g}\otimes{\bf g}$ (the {\em super
cocommutator}) such that

1)~~$\delta$ is a super one-cocycle, i.e.,\footnote{ Here
{\footnotesize $|X|(|Y|)$} indicates the gradings of $X$ and $Y$.
}
\begin{equation}
\delta([X,Y])=(ad_X\otimes I+I\otimes
ad_X)\delta(Y)-(-1)^{|X||Y|}(ad_Y\otimes I+I\otimes
ad_Y)\delta(X) \qquad \forall X,Y\in {\bf g},
\end{equation}

2) the dual map ${^t}{\delta}:{\bf g}^\ast\otimes {\bf g}^\ast \to
{\bf g}^\ast$ is a Lie superbracket on ${\bf g}^\ast$, i.e.,
\begin{equation}
(\xi\otimes\eta , \delta(X)) = ({^t}{\delta}(\xi\otimes\eta) , X)
= ([\xi,\eta]_\ast , X)    \qquad \forall X\in  {\bf g} ;\;\,
\xi,\eta\in{\bf g}^\ast.
\end{equation}
The Lie super-bialgebra thus defined is denoted by $({\bf g},{\bf
g}^\ast)$ or $({\bf g},\delta)$ \cite {{N.A},{ER1}}.

\smallskip

 {\it Definition}: A Lie super-bialgebra is {\em coboundary} if the
super cocommutator is a one-coboundary, i.e., if there exists  an
element $r\in{\bf g}\otimes{\bf g}$ such that
\begin{equation}
\delta(X) = (I\otimes ad_X + ad_X\otimes I)r        \qquad
\forall X\in {\bf g}.
\end{equation}

\smallskip

{\it Proposition}: Two coboundary Lie super-bialgebras $({\bf
g},{\bf g}^\ast)$ and $({\bf g}^\prime,{{\bf g}^\ast}^\prime)$
defined by $r\in \bf g \otimes \bf g$ and $r^\prime \in {\bf
g}^\prime \otimes {\bf g}^\prime$ are {\em isomorphic} if and only
if there is an isomorphism of Lie superalgebras $\alpha: \bf g
\longrightarrow {\bf g}^\prime$ such that $(\alpha\otimes\alpha)r
- r^\prime$ is ${\bf g}^\prime$ invariant \cite{RHR} i.e.,
\begin{equation}
(I\otimes ad_X + ad_X\otimes I)((\alpha\otimes\alpha)r -
r^\prime)=0 \qquad \forall X\in {\bf g}^\prime.
\end{equation}

\smallskip

 {\it Definition}: Coboundary Lie super-bialgebras
can be of two different types:\\

A.~if $r$ is a super skew-symmetric solution of the classical
Yang-Baxter equation (CYBE), i.e., $r^{ij} = -(-1)^{ij} r^{ji}$
and
\begin{equation}
[[r , r]] = 0,
\end{equation}
then the couboundary Lie super-bialgebra is said to be {\em
triangular}. In this case,  the  Schouten superbracket is defined
by\footnote {We note that $r$ has even Grassmann parity and
Grassmann parity of $r^{ij}$ comes from indices. For example, we
have$$ [r_{12},r_{13}]=(-1)^{i(k+l)+jl}\;r^{ij}r^{kl}\;[X_i,X_k]
\otimes X_j \otimes X_l
$$}

\begin{equation}
[[r , r]] = [r_{12}, r_{13}] + [r_{12} , r_{23}] + [r_{13} ,
r_{23}],
\end{equation}
and if we denote $r=r^{ij}X_i \otimes X_j$, then $r_{12}=
r^{ij}X_i \otimes X_j \otimes 1$, $r_{13}= r^{ij}X_i \otimes 1
\otimes X_j$ and $r_{23}= r^{ij}1 \otimes X_i \otimes X_j$. A
solution of the CYBE is often called a {\em classical r-matrix}.\\

B.~ if $r$ is a solution of CYBE such that $r_{12} + r_{21}$ is a
${\bf g}$ invariant element of ${\bf g}\otimes{\bf g}$, then the
coboundary Lie super-bialgebra is said to be {\em
quasi-triangular}. Moreover, if the super symmetric part of $r$ is
invertible, then $r$ is called {\em factorable}.

\smallskip
\goodbreak

Sometimes condition B can be replaced with the following:\\

$B'$.~ if $r$ is a super skew-symmetric solution of the modified
CYBE
\begin{equation}
[[r , r]] = \omega            \qquad \omega\in {\wedge}^3 {\bf g},
\end{equation}
then the coboundary Lie super-bialgebra is said to be
quasi-triangular \cite{N.A}.

We note that if ${\bf g}$ is a Lie super-bialgebra then ${\bf
g}^\ast $ is also a Lie super-bialgebra \cite{N.A}, but this is
not always true for the coboundary property.

\smallskip
\goodbreak

 {\it Definition}: Let  ${\bf g}$ be a
coboundary Lie super-bialgebra with super one-cocycle $\delta$ and
${\bf g}^\ast$ be also coboundary Lie super-bialgebra with the
super one-cocycle $\delta^{\ast}$
\begin{equation}
\forall\xi\in{\bf g}^\ast          \qquad\exists r^\ast\in{\bf
g}^\ast\otimes{\bf g}^\ast   \qquad \delta^{\ast}(\xi)=(I \otimes
ad_\xi+ ad_\xi\otimes I ) r^\ast,
\end{equation}
where $\delta^{\ast}: {\bf g}^\ast \longrightarrow {\bf
g}^\ast\otimes{\bf g}^\ast $. Then the pair $({\bf g},{\bf
g}^\ast)$ is called a bi-r-matrix Lie super-bialgebra \cite{RHR}
if the Lie superbracket $[.~ , ~. ]^\prime$ on ${\bf g}$ defined
via ${{^t}{\delta}^\ast}$ as
\begin{equation}
(\delta^\ast(\xi) , X \otimes Y) = (\xi , {{^t}{\delta}^\ast}(X
\otimes Y)) = (\xi , [X,Y]^\prime)    \qquad\forall X,Y\in{\bf g}
,\qquad \xi\in{\bf g}^\ast,
\end{equation}
is equivalent to the original one \cite{RHR} i.e.,
\begin{equation}
[X, Y]^\prime = S^{-1} [SX , SY]  \qquad \forall X,Y\in{\bf
g},\qquad S\in Aut({\bf g}).
\end{equation}

\smallskip

{\it Definition}:  A {\it Manin supertriple}   is a triple of Lie
superalgebras $(\cal{D} , {\bf g} , {\bf \tilde{g}})$ together
with a nondegenerate ad-invariant supersymmetric bilinear form
$<.~ , ~. >$ on $\cal{D}$, such that\hspace{2mm}

1.~~${\bf g}$ and ${\bf \tilde{g}}$ are Lie sub-superalgebras of
$\cal{D}$,\hspace{2mm}

2.~~$\cal{D} = {\bf g}\oplus{\bf \tilde{g}}$ as a supervector
space,\hspace{2mm}

3.~~${\bf g}$ and ${\bf \tilde{g}}$ are isotropic with respect to
$< .~, ~. >$,  i.e.,
\begin{equation}
<X_i , X_j> = <\tilde{X}^i , \tilde{X}^j> = 0, \hspace{10mm}
{\delta_i}^j=\;<X_i , \tilde{X}^j>\; = (-1)^{ij}<\tilde{X}^j,
X_i>=(-1)^{ij}{\delta^j}_i,
\end{equation}
where $\{X_i\}$ and $\{\tilde{X}^i\}$ are  bases of the respective
Lie superalgebras ${\bf g}$ and ${\bf \tilde{g}}$ \cite
{{N.A},{ER1}}. \\

We note that in the relation (15), ${\delta^j}_i$ is the ordinary
delta function. There is a one-to-one correspondence between Lie
super-bialgebra $({\bf g},{\bf g}^\ast)$ and Manin supertriple
$(\cal{D} , {\bf g} , {\bf \tilde{g}})$ with ${\bf \tilde{g}}=
{\bf g}^\ast$. If we choose the structure constants of Lie
superalgebras ${\bf g}$ and ${\bf \tilde{g}}$ as
\begin{equation}
[X_i , X_j] = {f^k}_{ij} X_k,\hspace{20mm} [\tilde{X}^i ,\tilde{ X}^j] ={{\tilde{f}}^{ij}}_{\; \; \: k} {\tilde{X}^k}, \\
\end{equation}
then ad-invariance of the bilinear form $< .~ , ~. >$ on $\cal{D}
= {\bf g}\oplus{\bf \tilde{g}}$ implies that \cite {ER1}
\begin{equation}
[X_i , \tilde{X}^j] =(-1)^j{\tilde{f}^{jk}}_{\; \; \; \:i} X_k
+(-1)^i {f^j}_{ki} \tilde{X}^k.
\end{equation}
Clearly, using Eqs. (6), (15) and (16) we obtain
\begin{equation}
\delta(X_i) = (-1)^{jk}{\tilde{f}^{jk}}_{\; \; \; \:i} X_j \otimes
X_k.
\end{equation}
We note that the appearance of $(-1)^{jk}$ in this relation is due
to the definition of natural inner product between ${\bf g}\otimes
{\bf g}$ and  ${\bf g}^\ast \otimes {\bf g}^\ast $ as $
(\tilde{X}^i \otimes \tilde{ X}^j ,X_k \otimes
X_l)=(-1)^{jk}{\delta^i}_k {\delta^j}_l$.  As a result, if we
apply this relation in the super one-cocycle condition (5), then
we obtain  super Jacobi identities (2) for the  dual Lie
superalgabra and the  mixed super Jacobi identities
\begin{equation}
{f^m}_{jk}{\tilde{f}^{il}}_{\; \; \; \; m}=
{f^i}_{mk}{\tilde{f}^{ml}}_{\; \; \; \; \; j} +
{f^l}_{jm}{\tilde{f}^{im}}_{\; \; \; \; \; k}+ (-1)^{jl}
{f^i}_{jm}{\tilde{f}^{ml}}_{\; \; \; \; \; k}+ (-1)^{ik}
{f^l}_{mk}{\tilde{f}^{im}}_{\; \; \; \; \; j}.
\end{equation}
This relation can also be obtained from super Jacobi identity of
$\cal{D}$.
 \vspace{4mm}
\section  {\large {\bf Two and three dimensional Lie superalgebras and
Lie super-bialgebras
}}

In \cite{ER1} we found and classified all two and three
dimensional Lie super-bialgebras for all two and three dimensional
indecomposable Lie superalgebras. The method of classification is
new and indeed it is improvement and generalization of the method
in \cite{JR}\footnote{We note that  \cite{Snobl} unfortunately
does not contain a standard, logical method for obtaining of low
dimensional Lie bialgebras.}. In this method,  we  use the adjoint
representation of super Jacobi identity (2) and mixed super Jacobi
identity (19) to find dual Lie superalgebras by direct
calculation; we then  use automorphism Lie supergroups of Lie
superalgebras to classify all nonisomorphic two and three
dimensional Lie super-bialgebras \cite{ER1} and \cite{ER4}.

Here for presentation of the notations that we will use and for
self consistence  of the paper, the list of two and three
dimensional indecomposable and decomposable Lie superalgebras
\cite{B} and their related Lie super-bialgebras \cite{ER1} and
\cite{ER4} are given in tables 1$-$4, respectively; note that as
we use DeWitt notation and standard basis here, the structure
constants $C^B_{FF}$ must be pure imaginary. In this section, we
use the classification of two and three dimensional Lie
superalgebras listed in \cite{B}. In that classification, Lie
superalgebras are divided into two types: trivial and nontrivial
Lie superalgebras for which the commutations of fermion-fermion is
zero or nonzero, respectively. The results have been presented in
table $1$. As the tables $(m, n-m)$ indicate, the Lie
superalgebras have $m, \{X_1,...,X_m\}$ bosonic and $n-m,
\{X_{m+1},...,X_n\}$ fermionic generators. For the labeling of the
trivial Lie superalgebras, the letters A, B, C with integral
superscripts i and real subscripts p, denote the equivalence
classes of Lie superalgebras of dimension d, where d= 1 for A, d=2
for B, d=3 for C. The superscript i and real subscripts p denote
the number of non isomorphic Lie superalgebras and the Lie
superalgebras parameters, respectively.  For the nontrivial Lie
superalgebras, we add to the bracketed symbol to the corresponding
trivial Lie superalgebra, where necessary, an integral superscript
and a real subscript.

\vspace{8mm}

{\bf Table 1} : {\small Two and three dimensional
decomposable\footnote{ Note that decomposable Lie
sueralgebras are as follows:\\
$B+A_{1,1}=B \oplus A_{1,1},\; (2A_{1,1}+A)=(A_{1,1}+A) \oplus
A_{1,0},\; C^1_0=C^1_{p=0},\;C^2_0=C^2_{p=0}=B \oplus A_{0,1}$ and
$(A_{1,1}+2A)^0=(A_{1,1}+A) \oplus A_{0,1}$. } and indecomposable
Lie superalgebras.\footnote{ The Lie superalgebra $A$ is  one
dimensional Abelian Lie superalgebra with one fermionic generator
where Lie superalgebra $A_{1,1}$ is its bosonization.
Furthermore, $(C^1_{\frac{1}{2}})$ is different
from $C^1_p$ and we show the latter by $C^1_{p=\frac{1}{2}}$.   }}\\
    \begin{tabular}{l l l l  l l p{25mm} }
    \hline\hline
   \vspace{-1mm}
{\footnotesize Type }& {\footnotesize ${\bf g}$ }& {\footnotesize
Bosonic } & {\footnotesize Fermionic }&{\footnotesize Non zero
(Anti) Commutation }&{\footnotesize Comments} \\

& & {\footnotesize  basis} & {\footnotesize basis}&{\footnotesize
relations}& \\\hline

{\footnotesize $(1,1)$}&{\footnotesize $B$}& {\footnotesize
$X_1$}& {\footnotesize $X_2$}&
{\footnotesize $[X_1,X_2]=X_2$}&{\footnotesize Trivial} \\

\vspace{1mm}

& {\footnotesize $(A_{1,1}+A)$}& {\footnotesize $X_1$}&
{\footnotesize $X_2$} &{\footnotesize
$\{X_2,X_2\}=X_1$}&{\footnotesize Nontrivial}
\\

\vspace{1mm}

{\footnotesize  $(2,1)$} & {\footnotesize  $B+A_{1,1}$}&
{\footnotesize  $X_1, X_2$}& {\footnotesize  $X_3$}&
{\footnotesize  $[X_1,X_3]=X_3$}& {\footnotesize Solvable, Trivial}\\

\vspace{1mm}

&{\footnotesize  $(2A_{1,1}+A)$} & {\footnotesize $X_1, X_2$} &
{\footnotesize $X_3$}&{\footnotesize
$\{X_3,X_3\}=X_1 $}&{\footnotesize   Nilpotent, Nontrivial}\\

\vspace{1mm}

&{\footnotesize  $C^1_0$}& {\footnotesize $X_1, X_2$ }&
{\footnotesize $X_3$}&
{\footnotesize $[X_1,X_2]=X_2$}& {\footnotesize Solvable,  Trivial}\\

\vspace{1mm}

& {\footnotesize ${C^1_p}$} & {\footnotesize $X_1,X_2$} &
{\footnotesize  $X_3$}&{\footnotesize
$[X_1,X_2]=X_2,\; [X_1,X_3]=pX_3 $}&{\footnotesize   $p\neq0$, Trivial } \\

\vspace{1mm}

&{\footnotesize  $(C^1_{\frac{1}{2}})$ } & {\footnotesize $X_1,X_2$}
& {\footnotesize  $X_3$}&{\footnotesize $[X_1,X_2]=X_2,\;
[X_1,X_3]=\frac{1}{2}X_3,\;\{X_3,X_3\}=X_2 $}&{\footnotesize
Nontrivial}\\

\vspace{1mm}

{\footnotesize $(1,2)$} & {\footnotesize $C^2_0$} &
{\footnotesize $X_1$} & {\footnotesize $ X_2, X_3$}&{\footnotesize
$[X_1,X_2]=X_2$}& {\scriptsize  Solvable, Trivial}\\

\vspace{1mm}

&{\footnotesize $(A_{1,1}+2A)^0$} & {\footnotesize $X_1$} &
{\footnotesize $ X_2, X_3$}&{\footnotesize
$\{X_2,X_2\}=X_1$}& {\footnotesize  Nilpotent, Nontrivial}\\

\vspace{1mm}

&{\footnotesize   $ C^2_p$} & {\footnotesize $X_1$} &
{\footnotesize  $X_2,X_3$}& {\footnotesize
$[X_1,X_2]=X_2, \;[X_1,X_3]=pX_3$ }&{\footnotesize  $0<|p|\leq 1$, Trivial } \\

\vspace{1mm}

&{\footnotesize $C^3$} & {\footnotesize  $X_1$} & {\footnotesize
$X_2,X_3$}  &{\footnotesize  $[X_1,X_3]=X_2
$}&{\footnotesize Nilpotent, Trivial  } \\

\vspace{1mm}

 &{\footnotesize  $C^4$} & {\footnotesize
$X_1$} & {\footnotesize  $X_2,X_3$}&{\footnotesize
$[X_1,X_2]=X_2,\; [X_1,X_3]=X_2+X_3$}
&{\footnotesize   Trivial }\\

\vspace{1mm}

&{\footnotesize $C^5_p$} & {\footnotesize  $X_1$} & {\footnotesize
$X_2,X_3$}&{\footnotesize $[X_1,X_2]=pX_2-X_3,\;
[X_1,X_3]=X_2+pX_3$ }&{\footnotesize   $p\geq0$, Trivial }
\\

\vspace{1mm}

&{\footnotesize $(A_{1,1}+2A)^1$}& {\footnotesize $X_1$} &
{\footnotesize $X_2,X_3$}&{\footnotesize $\{X_2,X_2\}=X_1,\;
\{X_3,X_3\}=X_1 $} &
{\footnotesize Nilpotent, Nontrivial} \\

\vspace{1mm}

& {\footnotesize $(A_{1,1}+2A)^2$}& {\footnotesize $X_1$} &
{\footnotesize  $X_2,X_3$}& {\footnotesize  $\{X_2,X_2\}=X_1,
\;\{X_3,X_3\}=-X_1 $}&{\footnotesize  Nilpotent, Nontrivial }
\smallskip \\
\hline\hline
\end{tabular}

\vspace{5mm}

 Note that some of these low dimensional Lie
superalgebras are Lie subsuperalgebras of the classical Lie
superalgebras as follows:
$$
~~~B \subset gl(1|1), ~osp(1|2), ~~~~~~~~~(A_{1,1}+A)\subset
osp(1|2),~~~~~~~~~B+A_{1,1}\subset gl(1|1),~~~~~~C^1_0 \subset
osp(1|2),
$$
\vspace{-4mm}
$$
~~~C^1_p\subset osp(1|2),~~~~~~~~~~~(C^1_{\frac{1}{2}}) \subset
osp(1|2), e(2),~~~~~~~~(A_{1,1}+2A)^0 \subset
e(2).~~~~~~~~~~~~~~~~~~~~~~~~~~~~~~~~~~~
$$

\newpage
{\small {\bf Table 2}}: {\small
Three dimensional  Lie super-bialgebras of the type (2 ,1).}\\
    \begin{tabular}{l l l l  p{0.15mm} }
    \hline\hline
{\footnotesize ${\bf g}$ }& {\footnotesize $\tilde{\bf g}$}
&{\footnotesize Non-zero (anti) commutation relations of
$\tilde{\bf g}$}& {\footnotesize Comments} \\ \hline

\vspace{2mm}

{\footnotesize $(2A_{1,1}+A) $}&{\footnotesize $I_{(2 , 1)} $}&
\\

\vspace{1mm}

{\footnotesize $B+A_{1,1}$}&{\footnotesize$I_{(2 , 1)} $}& &\\

\vspace{1mm}

&{\footnotesize $B+A_{1,1}|.i$}&{\footnotesize $[{\tilde X}^2,{\tilde X}^3]={\tilde X}^3$} & &\\

\vspace{1mm}

 &{\footnotesize $(2A_{1,1}+A)$} &{\footnotesize $\{{\tilde X}^3,{\tilde X}^3\}={\tilde X}^1$}
 &\\
\vspace{2mm}

 &{\footnotesize $(2A_{1,1}+A).i$} &{\footnotesize $\{{\tilde X}^3,{\tilde
X}^3\}=-{\tilde X}^1$} &\\

\vspace{1mm}

{\footnotesize $C^1_p$}&{\footnotesize $I_{(2 , 1)} $}& &{\footnotesize $p\in\Re$ }&\\

\vspace{1mm}

&{\footnotesize $(2A_{1,1}+A)$}&{\footnotesize $\{{\tilde X}^3,{\tilde X}^3\}={\tilde X}^1$} &{\footnotesize $p\in\Re$ }\\

\vspace{1mm}

&{\footnotesize $(2A_{1,1}+A).i$}& {\footnotesize $\{{\tilde
X}^3,{\tilde X}^3\}=-{\tilde X}^1$}
&{\footnotesize $p\in\Re$}\\

\vspace{1mm}

&{\footnotesize $(2A_{1,1}+A).ii$}& {\footnotesize $\{{\tilde
X}^3,{\tilde X}^3\}={\tilde X}^2$}&{\footnotesize
$p=\frac{1}{2}$}\\

\vspace{2mm}

&{\footnotesize $C^1_{-p}.i$ }&{\footnotesize $[{\tilde
X}^1,{\tilde X}^2]={\tilde X}^1,\;\;\;[{\tilde X}^2,{\tilde
X}^3]=p{\tilde X}^3$} &{\footnotesize $p\in\Re$}
\\

\vspace{2mm}

{\footnotesize $C^1_0$}&{\footnotesize
$C^1_{0,k}$}&{\footnotesize $[{\tilde X}^1,{\tilde X}^2]=k{\tilde
X}^2$}&{\footnotesize$k\in\Re-\{0\}$}
\\

\vspace{1mm}

{\footnotesize $(C^1_{\frac{1}{2}})$}&{\footnotesize $I_{(2 , 1)} $}&\\

\vspace{1mm}

&{\footnotesize
$C^1_{p}.i{_{|_{p=-\frac{1}{2}}}}$}&{\footnotesize$[{\tilde
X}^1,{\tilde X}^2]= {\tilde X}^1,\;\;\;[{\tilde X}^2,{\tilde
X}^3]=\frac{1}{2} {\tilde X}^3$}&
 \\
\vspace{1mm}

&{\footnotesize
$C^1_{p}.ii{_{|_{p=-\frac{1}{2}}}}$}&{\footnotesize$[{\tilde
X}^1,{\tilde X}^2]=-{\tilde X}^1,\;\;\;[{\tilde X}^2,{\tilde
X}^3]=-\frac{1}{2} {\tilde X}^3$}&\\

\vspace{1mm}

&{\footnotesize $(C^1_{\frac{1}{2}}).i$ }&{\footnotesize $[{\tilde
X}^1,{\tilde X}^2]={\tilde X}^1,\;\;\;[{\tilde X}^2,{\tilde
X}^3]=-\frac{1}{2} {\tilde X}^3,\;\;\; \{{\tilde X}^3,{\tilde
X}^3\}={\tilde X}^1$}&
\\

\vspace{1mm}

&{\footnotesize $(C^1_{\frac{1}{2}}).ii$ }&{\footnotesize $[{\tilde
X}^1,{\tilde X}^2]=-{\tilde X}^1,\;\;\;[{\tilde X}^2,{\tilde
X}^3]=\frac{1}{2} {\tilde X}^3,\;\;\; \{{\tilde X}^3,{\tilde
X}^3\}=-{\tilde X}^1$}&
\\

\vspace{1mm}

&{\footnotesize $({{C^1_{\frac{1}{2},k}}})$}&{\footnotesize
$[{\tilde X}^1,{\tilde X}^2]=k{\tilde X}^2,\;[{\tilde
X}^1,{\tilde X}^3]=\frac{k}{2}{\tilde X}^3,\;\{{\tilde
X}^3,{\tilde X}^3\}=k{\tilde X}^2$}&{\footnotesize $k\in
{\Re-\{0\}}$}
\smallskip \\
\hline\hline
\end{tabular}

 \vspace{12mm}

\vspace{6mm}

{\small {\bf Table 3}} : {\small Two dimensional
Lie super-bialgebras of the type $(1 ,1)$.}\\
    \begin{tabular}{l l l l p{5mm} }
    \hline\hline
{\footnotesize ${\bf g}$ }& \hspace{1cm}{\footnotesize
$\tilde{\bf g}$
}&\hspace{2cm}{\footnotesize (Anti) Commutation relations of $\tilde{\bf g}$}\\
\hline

\vspace{2mm}

{\footnotesize $(A_{1,1}+A)$} &\hspace{0.9cm} {\footnotesize
$I_{(1,1)}$}&
\\

\vspace{1mm}

{\footnotesize $B$} & \hspace{1cm}{\footnotesize $I_{(1,1)}$}
\\

\vspace{1mm}

&\hspace{1cm}{\footnotesize $(A_{1,1}+A)$} &\hspace{2cm}
{\footnotesize $\{{\tilde X}^2,{\tilde X}^2\}={\tilde X}^1$}&
\\

\vspace{-1mm}

&\hspace{1cm}{\footnotesize $(A_{1,1}+A).i$} &
\hspace{2cm}{\footnotesize $\;\{{\tilde X}^2,{\tilde X}^2\}=-
{\tilde X}^1$}&
\smallskip \\
\hline\hline
\end{tabular}

\newpage

{\small {\bf Table 4}}: {\small Three dimensional Lie
super-bialgebras of the type (1 ,2). where $\epsilon=
\pm 1$. }\\
   \begin{tabular}{l l l  p{0.5mm} }
    \hline\hline
{\footnotesize ${\bf g}$ }&  {\footnotesize $\tilde{\bf g}$} &
{\footnotesize Comments }\\ \hline \vspace{1mm}
{\footnotesize $C^2_1$}&$I_{(1 , 2)} $&\\
\vspace{1mm}
 &{\footnotesize $(A_{1,1}+2A)^0_{0,0,\epsilon }$ } & &\\
\vspace{1mm}

&{\footnotesize $(A_{1,1}+2A)^1_{\epsilon,0,\epsilon}$ } & & \\
\vspace{2mm}

&{\footnotesize  $(A_{1,1}+2A)^2_{\epsilon,0,-\epsilon}$} & &
\\

\vspace{1mm}
{\footnotesize $C^2_p$}&{\footnotesize  $I_{(1 , 2)} $}&\\
\vspace{1mm}

{\footnotesize $-1 \leq p <1$ }& {\footnotesize
$(A_{1,1}+2A)^0_{\epsilon ,
0,0}\;,\;\;(A_{1,1}+2A)^0_{0,0,\epsilon}\;,\;\;(A_{1,1}+2A)^0_{\epsilon,\epsilon,\epsilon}$}
&& \\
\vspace{1mm}

&{\footnotesize $(A_{1,1}+2A)^1_{\epsilon , k,\epsilon }$}
 &{\footnotesize $ -1<k<1$}& \\

\vspace{1mm}

&{\footnotesize $(A_{1,1}+2A)^2_{0 , 1,0
}\;,\;\;(A_{1,1}+2A)^2_{\epsilon,1,0}\;,\;\;(A_{1,1}+2A)^2_{0,1,\epsilon}\;,\;\;
(A_{1,1}+2A)^2_{\epsilon ,k,-\epsilon }$ }&{\footnotesize $k
\in\Re$}&
\\

\vspace{1mm}
{\footnotesize $C^3$}&{\footnotesize $I_{(1 , 2)} $}&\\
\vspace{1mm}

&{\footnotesize $(A_{1,1}+2A)^0_{1 ,0,0}\;,\;\;(A_{1,1}+2A)^0_{0,0,1}$ }& \\
\vspace{1mm}

&{\footnotesize $(A_{1,1}+2A)^1_{\epsilon , 0,\epsilon }$}
 && \\

\vspace{1mm}

& {\footnotesize $(A_{1,1}+2A)^2_{0 ,\epsilon,0
}\;,\;\;(A_{1,1}+2A)^2_{\epsilon ,0,-\epsilon}$}
 && \\

\vspace{1mm}
{\footnotesize $C^4$}& {\footnotesize $I_{(1 , 2)} $}&\\
\vspace{1mm}

& {\footnotesize $(A_{1,1}+2A)^0_{\epsilon ,0,0}\;,\;\;(A_{1,1}+2A)^0_{0,0,\epsilon}$ }&&\\
\vspace{1mm}

& {\footnotesize $(A_{1,1}+2A)^1_{k ,
0,1}\;,\;\;(A_{1,1}+2A)^1_{s , 0,-1}$}
 &{\footnotesize $0<k,\;\;s<0$}& \\

\vspace{1mm}

&{\footnotesize $(A_{1,1}+2A)^2_{0 ,\epsilon,0
}\;,\;\;(A_{1,1}+2A)^2_{k ,0,1}\;,\;\;(A_{1,1}+2A)^2_{s,0,-1}$}
 &{\footnotesize $k<0,\;\;0<s$}\\

\vspace{1mm}
{\footnotesize $C^5_p$}&{\footnotesize $I_{(1 , 2)} $}&\\
\vspace{1mm}

{\footnotesize $p\geq 0$}&{\footnotesize $(A_{1,1}+2A)^0_{0,0,\epsilon}$ }&&\\
\vspace{1mm}

&{\footnotesize $(A_{1,1}+2A)^1_{k , 0,1}\;,\;\;(A_{1,1}+2A)^1_{s
, 0,-1}$} &{\footnotesize $0<k,\;\;s<0$}& \\

\vspace{2mm}

& {\footnotesize  $(A_{1,1}+2A)^2_{k
,0,1}\;,\;\;(A_{1,1}+2A)^2_{s,0,-1}$}
 &{\footnotesize $ k<0,\;\;0<s$}\smallskip\smallskip\\

\vspace{1mm}

{\footnotesize $(A_{1,1}+2A)^0$}&{\footnotesize  $I_{(1 , 2)} $}&\\

\vspace{1mm}

{\footnotesize $(A_{1,1}+2A)^1$}& {\footnotesize $I_{(1 , 2)} $}&\\

\vspace{1mm}

{\footnotesize $(A_{1,1}+2A)^2$}&{\footnotesize  $I_{(1 , 2)} $}& \smallskip \\
\hline\hline
 \end{tabular}

\vspace{2mm}

Note that in table 4, all of the dual Lie superalgebras are non isomorphic. By considering
their subscripts indices, these  dual Lie superalgebras can be represented by the
notation: ${\tilde {\cal G}}_{\alpha\beta\gamma}\;=\;(A_{1,1}+2A)^0_{\alpha ,\beta,\gamma }\;,\;(A_{1,1}+2A)^1_{\alpha
,\beta,\gamma }$ and $(A_{1,1}+2A)^2_{\alpha ,\beta,\gamma }$, where we have the following
anticommutation relations:
\begin{equation}
\{{\tilde X}^2,{\tilde X}^2\}=\alpha {\tilde X}^1 ,\quad
\{{\tilde X}^2,{\tilde X}^3\}= \beta {\tilde X}^1 ,\quad
\{{\tilde X}^3,{\tilde X}^3\}=\gamma {\tilde X}^1,
\end{equation}
where $ \alpha, \beta, \gamma = 0, 1, -1, \epsilon, k, s.$

\vspace{7mm}

\section  {\large {\bf Two and three dimensional coboundary Lie super-bialgebras
}}

In this section, we determine how many of $74$ of two and three
dimensional Lie super-bialgebras of  tables 2$-$4  are coboundary?
Note that here we work in  nonstandard basis, so we omit
coefficient $i=\sqrt{-1}$ from all anticommutation relations for
Lie superalgebras and Lie super-bialgebras of \cite{ER1} and
\cite{ER4}. In this way, we must find $r=r^{ij}X_i \otimes X_j
\in {\bf g}\otimes{\bf g}$ such that the super cocommutator of Lie
super-bialgebras can be written as (7). Using Eqs. (7), (16) and
(18), we have
\begin{equation}
{\tilde{\cal Y}}_i = {{\cal X}_i}^{st} r + (-1)^l\;r {\cal X}_i,
\end{equation}
where index $l$ corresponds to the row of matrix ${\cal X}_i$ and
${({{\cal X}}_i)_l}^k= -{{f}_{il}}^{\;k}$ and ${({\tilde{\cal
Y}}_i)^{jk}}= -{\tilde{f}^{jk}}_{\; \; \; \:i}$ are adjoint
representations for the Lie superalgebras ${\bf g}$ and
$\tilde{\bf g}$, respectively, [here superscript $st$ stands for
supertranspose].  Now using  the above relations, we can find the
r-matrix of the Lie super-bialgebras. In this manner, we
determine which of presented Lie super-bialgebras in tables
2$-$4  are coboundary and obtain their r-matrices. We also
perform this work for the dual Lie super-bialgebras $(\tilde{\bf
g}, {\bf g})$  using the following equations as (21)
\begin{equation}
{\cal Y}^i = ({\tilde{\cal X}}^i)^{st} \tilde{r} +
(-1)^l\;\tilde{r} {\tilde{\cal X}}^i,
\end{equation}
where as above, ${({\tilde{\cal X}}^i)^j}_l= -{\tilde{f}^{ij}}_{\;
\; \; \:l}$ and ${({{\cal Y}}^i)_{jk}}= -{{f}_{jk}}^{\; i}$. The
results are summarized in tables 5$-$7. Note that for
determining  Schouten superbrackets on Poisson-Lie supergroups,
information on the type of Lie super-bialgebras (triangular or
quasi-triangular) is important. As a result,  we classify all
types of two and three dimensional coboundary Lie
super-bialgebras. Two points should be highlighted concerning
these  tables. First, we have listed coboundary Lie
super-bialgebras $({\bf g}, \tilde{\bf g})$ with coboundary duals
$({\bf {\tilde g}}, {\bf g})$ in all tables. Since such
structures can be specified (up to automorphism) by pairs of
r-matrices, then it is natural to call them bi-r-matrix
super-bialgebras (b-r-sb) \cite{RHR}. Here, we give complete list
of two and three dimensional coboundary and  b-r-sb. Secondly, as
it is clearly seen, we have considered super skew-symmetric
r-matrix solutions in tables 5$-$7. Of course there are other
solutions for some Lie super-bialgebras of these tables. We have
also listed these solutions in those tables.

\vspace{9mm}
{ {\bf Table 5}}: ~{\small Two dimensional coboundary and
bi-r-matrix Lie super-bialgebras of the type (1,1).\smallskip\\
\smallskip
\begin{tabular}{l l l l l l  p{20mm} }
    \hline\hline
\smallskip

{\footnotesize $({\bf g}, \tilde{\bf g})$ }&{\footnotesize  $r$}
&{\footnotesize $[[r , r]]$}&{\footnotesize ${\tilde
r}$}&{\footnotesize $[[\tilde r , \tilde r]]$}
&{\footnotesize Comments}\smallskip \\
\hline
\smallskip

\vspace{2mm}

{\footnotesize $((A_{1,1}+A),I_{(1,1)})$}&
{\footnotesize  $aX_1 \otimes X_1$} &{\footnotesize $0$} & & &{\footnotesize  $a \in \Re$}\\

\vspace{2mm}

{\footnotesize $(B,(A_{1,1}+A))$}& {\footnotesize $-\frac{1}{4}
X_2 \wedge X_2$} &{\footnotesize $0$} &{\footnotesize $a{\tilde
X}^1 \otimes {\tilde X}^1+\frac{1}{2} {\tilde X}^2 \wedge {\tilde
X}^2 $}&{\footnotesize $-\frac{1}{2}{\tilde
X}^1\wedge {\tilde X}^2 \wedge {\tilde X}^2 $}&{\footnotesize $a \in\Re$}\\

\vspace{2mm}

{\footnotesize $(B,(A_{1,1}+A).i)$}&{\footnotesize  $ \frac{1}{4}
X_2 \wedge X_2$} &{\footnotesize $0$}&{\footnotesize $a{\tilde
X}^1 \otimes {\tilde X}^1-\frac{1}{2} {\tilde X}^2 \wedge {\tilde
X}^2 $}&{\footnotesize $ \frac{1}{2}{\tilde X}^1 \wedge {\tilde
X}^2 \wedge {\tilde X}^2 $}&{\footnotesize $a
\in\Re$}\\\hline\hline

\end{tabular}

\vspace{6mm}

Note that as shown in table 5 we have two bi-r-matrix Lie
super-bialgebras such that the Lie super-bialgebras
$(B,(A_{1,1}+A))$ and $(B,(A_{1,1}+A).i)$ are triangular while
their duals i.e., the Lie super-bialgebras $((A_{1,1}+A) , B)$ and
$((A_{1,1}+A).i , B)$ are quasi-triangular. Furthermore to obtain
super skew-symmetric  Poisson superbrackets for
$((A_{1,1}+A),I_{(1,1)})$ we must put  $a=0$ i.e., $r=0$, in this
case we have trivial solution.\\

Note that,  generally speaking, we have
$$
X_i \wedge X_j \wedge X_k=X_i \otimes X_j \otimes
X_k+(-1)^{i(j+k)} X_j \otimes X_k \otimes X_i+(-1)^{k(i+j)} X_k
\otimes X_i \otimes X_j \hspace{3cm}
$$
$$
-(-1)^{jk} X_i \otimes X_k \otimes X_j-(-1)^{ij} X_j \otimes X_i
\otimes X_k-(-1)^{ij+ik+jk} X_k \otimes X_j \otimes X_i.
$$ }

\newpage

{\bf Table 6}:~{\small Three dimensional coboundary and
bi-r-matrix Lie super-bialgebras of the type
$(2,1)$.\\\vspace{0.5mm} \hspace{19.5mm}where
$\zeta, \eta$ are a-numbers. }\\
 \begin{tabular}{l l l l p{0.5mm} }
  \hline\hline

{\footnotesize $({\bf g}, \tilde{\bf g})$ }&
&{\footnotesize Comments}\smallskip \\
\hline
\smallskip
\vspace{-1mm}
{\footnotesize $((2A_{1,1}+A) , I_{(2,1)})$}&{\footnotesize $r =
aX_1 \otimes X_1+ bX_1
\otimes X_2+c X_2 \otimes X_1+dX_2 \otimes X_2$}&{\footnotesize $a, b, c, d \in \Re$}\\

\vspace{2mm}

&{\footnotesize $[[r , r]]=0$}&\\
\vspace{-1mm}

{\footnotesize $(B+A_{1,1} , I_{(2,1)})$}& {\footnotesize
$r=aX_2 \otimes X_2$}&
{\footnotesize $a \in \Re$}\\

\vspace{2mm}

&{\footnotesize $[[r , r]]=0$}\\
\vspace{1mm}

{\footnotesize $(B+A_{1,1} , B+A_{1,1}|.i)$}&{\footnotesize
 $r = aX_2 \otimes X_2+X_1 \wedge X_2 $ }& {\footnotesize $a \in \Re$}\\

\vspace{1mm}

&{\footnotesize $[[r , r]]=0$}\\

\vspace{1mm}

&{\footnotesize ${\tilde r}=b {\tilde X}^1 \otimes {\tilde
X}^1-{\tilde X}^1
\wedge {\tilde X}^2$}&{\footnotesize $b \in\Re$}\\

\vspace{2mm}

&{\footnotesize $[[\tilde r , \tilde r]]=0$}\\

\vspace{1mm}

{\footnotesize $(B+A_{1,1} , (2A_{1,1}+A))$}& {\footnotesize
$r = aX_2 \otimes X_2-\frac{1}{4} X_3 \wedge X_3 $ }&{\footnotesize $a \in \Re$}\\

\vspace{1mm}

&{\footnotesize $[[r , r]]=0$}\\

\vspace{1mm}

&{\footnotesize ${\tilde r}= b {\tilde X}^1 \otimes {\tilde X}^1+c
{\tilde X}^1 \otimes {\tilde X}^2+ d{\tilde X}^2\otimes {\tilde
X}^1+e {\tilde X}^2 \otimes {\tilde X}^2+\frac{1}{2} {\tilde X}^3
\wedge
{\tilde X}^3$}&{\footnotesize $b, c, d, e \in\Re$}\\

\vspace{2mm}

&{\footnotesize $[[\tilde r , \tilde r]]=-\frac{1}{2} {\tilde X}^1
\wedge
{\tilde X}^3 \wedge {\tilde X}^3$}\\

\vspace{1mm}

{\footnotesize $(B+A_{1,1} , (2A_{1,1}+A).i)$}& {\footnotesize
$r= aX_2 \otimes X_2+\frac{1}{4} X_3 \wedge X_3 $}
&{\footnotesize $a \in \Re$}\\

\vspace{1.5mm}

&{\footnotesize $[[r , r]]=0$}\\

\vspace{1mm}

&{\footnotesize ${\tilde r}= b {\tilde X}^1 \otimes {\tilde X}^1+c
{\tilde X}^1 \otimes {\tilde X}^2+ d{\tilde X}^2\otimes {\tilde
X}^1+e {\tilde X}^2 \otimes {\tilde X}^2-\frac{1}{2} {\tilde X}^3
\wedge
{\tilde X}^3$}&{\footnotesize $b, c, d, e \in\Re$}\\

\vspace{2mm}

&{\footnotesize $[[\tilde r , \tilde r]]=\frac{1}{2} {\tilde X}^1
\wedge {\tilde
X}^3 \wedge {\tilde X}^3$}\\

\vspace{1mm}

{\footnotesize $(C^1_p , (2A_{1,1}+A))$}&{\footnotesize
 $r_1=-\frac{1}{4p}X_3 \wedge X_3$}&{\footnotesize $p\in \Re-\{0,1\}$}\\

\vspace{1.5mm}

&{\footnotesize $[[r_1 , r_1]]=0$}\\

\vspace{1mm}

&{\footnotesize $r_2=\zeta X_2 \otimes X_3+\eta X_3 \otimes
X_2-\frac{1}{4}X_3
\wedge X_3$}&{\footnotesize $p=1$}\\

\vspace{2mm}

&{\footnotesize $[[r_2 , r_2]]=0$}\\
\vspace{1mm}

{\footnotesize $(C^1_p , (2A_{1,1}+A).i)$}&{\footnotesize
 $r_1=\frac{1}{4p}X_3 \wedge X_3$}&{\footnotesize $p\in \Re-\{0,1\}$}\\

\vspace{1.5mm}

&{\footnotesize $[[r_1 , r_1]]=0$}\\

\vspace{1mm}

&{\footnotesize $r_2=\zeta X_2 \otimes X_3+\eta X_3 \otimes
X_2+\frac{1}{4}X_3
\wedge X_3$}&{\footnotesize $p=1$}\\

\vspace{2mm}

&{\footnotesize $[[r_2 , r_2]]=0$}\\
\vspace{1mm}

{\footnotesize $( C^1_{p} , C^1_{-p}.i)$}&{\footnotesize $r_1=
X_1 \wedge X_2+\frac{a}{2} X_3 \wedge X_3$}
&{\footnotesize $p=0,\;\;a\in \Re$ }\\

\vspace{2mm}

&{\footnotesize $[[r_1 , r_1]]=0$}\\

\vspace{1mm}

&{\footnotesize $r_2=X_1 \wedge X_2+\zeta X_2 \wedge X_3$ }&{\footnotesize $p=1$} \\

\vspace{2mm}

&{\footnotesize $[[r_2 , r_2]]=0$}\\

\vspace{1mm}

&{\footnotesize $r_3= X_1 \wedge X_2$ }&{\footnotesize $p\in \Re-\{0,1\}$} \\

\vspace{2mm}

&{\footnotesize $[[r_3 , r_3]]=0$}\\

\vspace{1mm}

&{\footnotesize ${\tilde r_1}=-{\tilde X}^1 \wedge {\tilde
X}^2+\frac{b}{2}
{\tilde X}^3 \wedge {\tilde X}^3$}&{\footnotesize $p=0,\;\;b\in \Re$}\\

\vspace{2mm}

&{\footnotesize $[[\tilde r_1 , \tilde r_1]]=0$}\\

\vspace{1mm}

&{\footnotesize ${\tilde r_2}=-{\tilde X}^1 \wedge {\tilde
X}^2+\zeta
{\tilde X}^1 \wedge {\tilde X}^3$}&{\footnotesize $p=-1$}\\

\vspace{2mm}

&{\footnotesize $[[\tilde r_2 , \tilde r_2]]=0$}\\

\vspace{1mm}

&{\footnotesize ${\tilde r_3}=-{\tilde X}^1 \wedge {\tilde X}^2 $}&
{\footnotesize $p\in \Re-\{0,-1\}$}\\

\vspace{2mm}

&{\footnotesize $[[\tilde r_3 , \tilde r_3]]= 0$}\\
\hline

\end{tabular}

\smallskip
\goodbreak
\vspace{1cm}
{\small{ { \bf Table 6}-Continued.}\footnote{Note that in the above
table, $(C^1_{\frac {1}{2},\epsilon=1})$ and $(C^1_{\frac
{1}{2},\epsilon=-1})$ denote the dual Lie superalgebras for
$(C^1_{\frac {1}{2}}.i)$ and $(C^1_{\frac {1}{2}}.ii)$, respectively.}}
\smallskip\\
\begin{tabular}{l l l l p{0.5mm}}
\hline \hline
{\footnotesize $({\bf g}, \tilde{\bf g})$ }&
&{\footnotesize Comments}\smallskip \\
\hline
\smallskip
\vspace{-1mm}
{\footnotesize $(C^1_p , I_{(2,1)})$}&\hspace{1cm}{\footnotesize
$r_1=\frac{a}{2}X_3 \wedge X_3$}
& {\footnotesize $p=0,\;\;a \in \Re$}\\

\vspace{1.5mm}

&\hspace{1cm}{\footnotesize $[[r_1 , r_1]]= 0$}\\

\vspace{1mm}

&\hspace{1cm} {\footnotesize $r_2=\zeta X_2 \otimes X_3+\eta X_3
\otimes X_2$}
& {\footnotesize $p=1$}\\

\vspace{2mm}

&\hspace{1cm}{\footnotesize  $[[r_2 , r_2]]=0$}\\
{\footnotesize $(
(C^1_{\frac{1}{2}}) , C^1_{p}.i{_{|_{p=-\frac{1}{2}}}})$}&\hspace{1cm}{\footnotesize
$r= X_1 \wedge X_2 $}
& \\

\vspace{1.5mm}

&\hspace{1cm}{\footnotesize $[[r , r]]= 0$}\\

\vspace{1mm}

&\hspace{1cm}{\footnotesize ${\tilde r}= -{\tilde X}^1 \wedge
{\tilde X}^2-\frac{1}{2}
{\tilde X}^3 \wedge {\tilde X}^3$}&\\

\vspace{2mm}

&\hspace{1cm}{\footnotesize  $[[\tilde r , \tilde r]]=\frac{1}{2}
{\tilde X}^1 \wedge
{\tilde X}^3 \wedge {\tilde X}^3$}\\

{\footnotesize $(
(C^1_{\frac{1}{2}}) , C^1_{p}.ii{_{|_{p=-\frac{1}{2}}}})$}&
\hspace{1cm}
{\footnotesize $r= -X_1 \wedge X_2 $ }& \\

\vspace{1.5mm}

&\hspace{1cm}{\footnotesize $[[r , r]]= 0$}\\

\vspace{1mm}

&\hspace{1cm} {\footnotesize ${\tilde r}= {\tilde X}^1 \wedge
{\tilde X}^2+\frac{1}{2}
{\tilde X}^3 \wedge {\tilde X}^3$}&\\

\vspace{2.5mm}

&\hspace{1cm} {\footnotesize $[[\tilde r , \tilde
r]]=-\frac{1}{2} {\tilde X}^1 \wedge {\tilde X}^3 \wedge {\tilde
X}^3$}\vspace{1mm}\\
{\footnotesize $(
(C^1_{\frac{1}{2}}) , (C^1_{\frac{1}{2},\epsilon}))$}& \hspace{1cm}
{\footnotesize $r= \epsilon X_1 \wedge X_2-\frac{\epsilon}{2} X_3
\wedge X_3$ }
& {\footnotesize $\epsilon=\pm 1$}\\

\vspace{1.5mm}

&\hspace{1cm}{\footnotesize $[[r , r]]= 0$}\\

\vspace{1mm}

&\hspace{1cm} {\footnotesize ${\tilde r}= -\frac{1}{\epsilon}
{\tilde X}^1 \wedge {\tilde X}^2+\frac{1}{2\epsilon}
{\tilde X}^3 \wedge {\tilde X}^3$}& {\footnotesize $\epsilon=\pm 1$}\\

\vspace{02.5mm}

&\hspace{1cm} {\footnotesize $[[\tilde r , \tilde r]]= 0$}
\\\hline\hline

\end{tabular}


\vspace{8mm}

In table 6 $((2A_{1,1}+A) , I_{(2,1)})$ is triangular Lie
super-bialgebras such that as shown in the next section to obtain
super skew-symmetric  Poisson superbrackets we must put
$a=d=0,\;b=-c$  so that its r-matrix has the form $r=b {X}_1
\wedge { X}_2$.

\smallskip

In the same way for triangular Lie super-bialgebra $(B+A_{1,1} ,
I_{(2,1)})$ we have $r=0$ and for bi-r-matrix Lie super-bialgebra
$(B+A_{1,1} , B+A_{1,1}|.i)$ with triangular type we have ${
r}={X}_1 \wedge {X}_2$ (i.e., $a=0$) and ${\tilde r}=-{\tilde X}^1
\wedge {\tilde X}^2$ (i.e., $b=0$) and for bi-r-matrix Lie
super-bialgebras $(B+A_{1,1} , (2A_{1,1}+A))$ and $(B+A_{1,1} ,
(2A_{1,1}+A).i)$ with triangular type from one hand  and
quasi-triangular type for their duals, we have $r=-\frac{1}{4}{
X}_3 \wedge { X}_3$ (i.e., $a=0$) and $r=\frac{1}{4}{ X}_3 \wedge
{ X}_3$ (i.e., $a=0$), respectively.

\smallskip

Note that the Lie super-bialgebras $(C^1_p, (2A_{1,1}+A))$ and
$(C^1_p, (2A_{1,1}+A).i)$ are triangular and have not coboundary
duals. Furthermore, for obtaining super skew-symmetric  Poisson
superbrackets we must put $r_2=\zeta { X}_2 \wedge { X}_3\mp
\frac{1}{4}{ X}_3 \wedge { X}_3$ (for $p=1$) i.e., $\eta=-\zeta$.
The Lie super-bialgebras $(C^1_p , C^1_{-p}.i)$ and their duals
for value $p=0\; (r_1$ and  ${\tilde r}_1),\; p=1 \;(r_2$ and ${
\tilde r}_{3_|{_{p=1}}}),\;p=-1\;({r}_{3_|{_{p=-1}}}$ and ${\tilde
r}_2)$ and $p\in\Re-\{-1,0,1\}$ $(r_3$ and $\tilde r_3)$ are
bi-r-matrix Lie super-bialgebras with triangular type.

\smallskip

The Lie super-bialgebras $(C^1_p , I_{(2,1)})$ are triangular for
values $p=0$ and $p=1$, furthermore, to obtain super
skew-symmetric super Poisson brackets we must put $r_2=\zeta {X}_2
\wedge { X}_3$ i.e., $\eta =-\zeta$. The $( (C^1_{\frac{1}{2}}) ,
C^1_{p}.i{_{|_{p=-\frac{1}{2}}}})$ and $( (C^1_{\frac{1}{2}}) ,
C^1_{p}.ii{_{|_{p=-\frac{1}{2}}}})$ are bi-r-matrix Lie
super-bialgebras with triangular type while their duals  are
quasi-triangular. The Lie super-bialgebras $( (C^1_{\frac{1}{2}})
, (C^1_{\frac{1}{2},\epsilon}))$  and their duals are bi-r-matrix
Lie super-bialgebras with triangular type.

\smallskip
\goodbreak

\newpage
{ {\bf Table 7}}:~{\small Three dimensional coboundary and
bi-r-matrix Lie super-bialgebras of the type
$(1,2)$.}\\\vspace{0.5mm} \hspace{20mm}where
$\zeta, \eta$ are a-numbers.\\
  \begin{tabular}{l l l l p{0.5mm} }
   \hline\hline

 {\footnotesize $({\bf g}, \tilde{\bf g})$ }&
&{\footnotesize Comments}\smallskip \\
\hline
\smallskip
\vspace{-1mm}
{\footnotesize  $(C^2_p , \cal{\tilde {G}}_{\alpha\beta\gamma}) $}&
{\footnotesize $r_1=-\frac{\alpha}{4} X_2 \wedge X_2-\beta X_2
\wedge X_3+\frac{b}{2} X_3 \wedge X_3
$}&{\footnotesize  $p=0;\;\;\gamma=0;\;\; b \in \Re$}\\

\vspace{1.2mm}

{\footnotesize $|p|\leq 1$ }&{\footnotesize  $[[r_1 , r_1]]=0$}&\\

\vspace{0.5mm}

&{\footnotesize  $r_2=-\frac{\alpha}{4} X_2 \wedge X_2+b X_2
\otimes X_3+c X_3 \otimes X_2+ \frac{\gamma}{4} X_3 \wedge
X_3$}&{\footnotesize
 $p=-1;\;\;\beta=0;\;\; b,c \in \Re$}\\

\vspace{1.2mm}

&{\footnotesize  $[[r_2 , r_2]]=0$}& \\

\vspace{0.5mm}

&{\footnotesize $r_3=-\frac{\alpha}{4} X_2 \wedge
X_2-\frac{\beta}{1+p} X_2 \wedge X_3-\frac{\gamma}{4p} X_3 \wedge
X_3$}&{\footnotesize
 $p\in (-1,1]-\{0\}$}\\
\vspace{1.2mm}

&{\footnotesize  $[[r_3 , r_3]]=0$} &\\

\vspace{0.5mm}

&{\footnotesize  $\tilde r_1= b {\tilde X}^1 \otimes {\tilde
X}^1+\zeta {\tilde X}^1 \otimes {\tilde X}^3+\eta {\tilde X}^3
\otimes {\tilde X}^1 +\frac{1}{2 \alpha} {\tilde X}^2 \wedge
{\tilde X}^2+ \frac{c}{2} {\tilde X}^3 \wedge {\tilde
X}^3$} & {\footnotesize  $p=0;\;\;\beta=\gamma =0;\;\;\alpha\neq 0;$}\\

\vspace{1mm}

&{\footnotesize  $[[\tilde r_1 , \tilde r_1]]= -\frac{1}{2 \alpha}
{\tilde X}^1 \wedge {\tilde X}^2 \wedge {\tilde X}^2
$}&{\footnotesize $ b, c \in \Re$}\\

\vspace{0.5mm}

&{\footnotesize  $\tilde r_2= b {\tilde X}^1 \otimes {\tilde
X}^1+\frac{1}{2\alpha } {\tilde X}^2 \wedge {\tilde
X}^2+\frac{p}{2 \gamma} {\tilde X}^3 \wedge {\tilde X}^3
$} &{\footnotesize  $p\in [-1,1]-\{0\};\;\;\;\; b \in \Re$}\\

\vspace{2mm}

&{\footnotesize  $[[\tilde r_2 , \tilde r_2]]=-\frac{1}{2 \alpha}
{\tilde X}^1 \wedge {\tilde X}^2 \wedge {\tilde X}^2-\frac{p^2}{2
\gamma} {\tilde X}^1 \wedge {\tilde X}^3 \wedge {\tilde X}^3
$}&{\footnotesize $\beta= 0;\;\;\alpha,
\gamma \neq 0$}\\
\vspace{0.5mm}

{\footnotesize $(C^3 , I_{(1,2)}) $}&{\footnotesize $r= \zeta X_1
\wedge X_2+\frac{c}{2} X_2 \wedge X_2+d (X_2 \otimes X_3- X_3
\otimes
X_2)$}& {\footnotesize $ c, d \in \Re$}\\

\vspace{2mm}

&{\footnotesize $[[r , r]]=0$}\\
\vspace{0.5mm}
{\footnotesize $(C^3 , (A_{1,1}+2A)^0_{1,0,0})
$}&{\footnotesize $r= \frac{b}{2} X_2 \wedge X_2+c X_2 \otimes
X_3-(1+c) X_3 \otimes X_2
$}& {\footnotesize $b, c \in \Re$}\\

\vspace{1mm}

&{\footnotesize $[[r , r]]= 0$}\\

\vspace{0.5mm}

& {\footnotesize $\tilde r= m{\tilde X}^1 \otimes {\tilde
X}^1+\zeta({\tilde X}^1 \otimes {\tilde X}^3+ {\tilde X}^3
\otimes {\tilde X}^1) +{\tilde X}^2 \wedge {\tilde
X}^3+\frac{n}{2} {\tilde X}^3 \wedge {\tilde X}^3$}
& {\footnotesize $ m, n \in \Re$}\\

\vspace{2mm}

&{\footnotesize $[[\tilde r , \tilde r]]= -\frac{1}{2} {\tilde
X}^1 \wedge {\tilde X}^3\wedge {\tilde
X}^3$} &\\
\vspace{0.5mm}
{\footnotesize $(C^3 , (A_{1,1}+2A)^2_{0,\epsilon,0})
$}&{\footnotesize  $r=\zeta X_1 \wedge X_2+\frac{c}{2} X_2 \wedge
X_2+d (X_2 \otimes X_3- X_3 \otimes
X_2)-\frac{\epsilon}{2} X_3 \wedge X_3 $}&{\footnotesize  $ c,d \in \Re$}\\

\vspace{1mm}

&{\footnotesize $[[r , r]]=\epsilon \zeta X_2 \wedge X_3 \wedge
X_2$}&\\

\vspace{0.5mm}

& {\footnotesize $\tilde r= m {\tilde X}^1 \otimes {\tilde
X}^1+\eta ({\tilde X}^1 \otimes {\tilde X}^3+ {\tilde X}^3
\otimes {\tilde X}^1) +\frac{1}{2 \epsilon} {\tilde X}^3 \wedge
{\tilde X}^3$}
& {\footnotesize $m \in \Re$}\\

\vspace{2mm}

&{\footnotesize $[[\tilde r , \tilde r]]= 0$} &\\

\vspace{0.5mm}

{\footnotesize $(C^4 , \cal{\tilde {G}}_{\alpha\beta\gamma})
$}&{\footnotesize $r= \frac{2(\beta-\alpha)-\gamma}{8}X_2 \wedge
X_2+\frac{\gamma-2\beta}{4} X_2\wedge
X_3-\frac{\gamma}{4} X_3\wedge X_3$}&\\

\vspace{1mm}

&{\footnotesize $[[r , r]]= 0$}\\

\vspace{0.5mm}

&{\footnotesize ${\tilde r}= a {\tilde X}^1\otimes {\tilde
X}^1+\frac{1}{\beta} {\tilde X}^2 \wedge {\tilde
X}^3+\frac{1}{2\beta}{\tilde X}^3 \wedge {\tilde X}^3$}&
{\footnotesize $\alpha=\gamma=0,\;\beta\neq 0;\;\;a\in\Re$}\\

\vspace{2mm}

&{\footnotesize $[[\tilde r , \tilde r]]= -\frac{1}{\beta} {\tilde
X}^1 \wedge {\tilde X}^2 \wedge {\tilde
X}^3+\frac{1}{\beta}{\tilde X}^3 \wedge
 {\tilde X}^1 \wedge {\tilde X}^3$}\\
\vspace{0.5mm}

{\footnotesize $(C^5_p , {\cal{\tilde {G}}}_{\alpha 0 \gamma}),\;\;
$}{\footnotesize $p\geq 0$}&{\footnotesize $r_1=\frac{a}{2} X_2
\wedge X_2+\frac{a}{2} X_3\wedge X_3+b X_2\otimes
X_3+(\gamma-b)X_3\otimes X_2$}&{\footnotesize  $p=0;\;\;\alpha
+\gamma =0;\;\;a,b\in\Re$}\\

\vspace{1mm}

&{\footnotesize $[[r_1 , r_1]]=0$}&\\

\vspace{0.5mm}

&{\footnotesize $r_2=-\frac{2\alpha p^2+\alpha+\gamma}{8p(1+p^2)}
X_2\wedge X_2+\frac{\gamma-\alpha}{4p(1+p^2)} X_2\wedge
X_3-\frac{2\gamma p^2+\alpha+\gamma}{8p(1+p^2)} X_3\wedge
X_3$}&{\footnotesize $p > 0$}\\

\vspace{1mm}

&{\footnotesize $[[r_2 , r_2]]=0$}&\\

\vspace{0.5mm}

&{\footnotesize ${\tilde r}= a {\tilde X}^1\otimes {\tilde
X}^1-\frac{p}{2\gamma}({\tilde X}^2 \wedge {\tilde X}^2-{\tilde
X}^3 \wedge {\tilde X}^3)-\frac{1}{\gamma}{\tilde X}^2
\wedge {\tilde X}^3$}&{\footnotesize  $p\geq 0;\;\;\alpha +\gamma =0;\;\;a\in\Re;$}\\

\vspace{2mm}

&{\footnotesize $[[\tilde r , \tilde r]]= \frac{ p^2-1}{2\gamma}
({\tilde X}^1 \wedge {\tilde X}^2 \wedge {\tilde X}^2 -{\tilde
X}^1 \wedge {\tilde X}^3 \wedge {\tilde X}^3)+\frac{2p}{\gamma}
{\tilde X}^1 \wedge {\tilde X}^2 \wedge {\tilde X}^3
$}&{\footnotesize $\gamma \neq 0$}\\
\vspace{0.5mm}

{\footnotesize  $((A_{1,1}+2A)^0 , I_{(1,2)})$} &{\footnotesize
$r= aX_1 \otimes X_1+\zeta X_1 \otimes X_3+\eta X_3 \otimes X_1
+\frac{d}{2}X_3 \wedge X_3$}&{\footnotesize  $a, d \in \Re$}\\

\vspace{2mm}

&{\footnotesize $[[r , r]]= 0$}\\
\vspace{0.5mm}

{\footnotesize  $((A_{1,1}+2A)^1 , I_{(1,2)})$}&{\footnotesize
$r= aX_1 \otimes X_1$}
&{\footnotesize  $a \in \Re$}\\
\vspace{2mm}

&{\footnotesize $[[r , r]]=0$}\\
\vspace{0.5mm}

{\footnotesize  $((A_{1,1}+2A)^2 , I_{(1,2)})$}&{\footnotesize
$r=aX_1 \otimes X_1+\zeta [X_1 \otimes X_2+X_2 \otimes X_1 \pm
X_1 \otimes X_3 \pm
X_3 \otimes X_1]$}&{\footnotesize  $a \in \Re$}\\
\vspace{1mm}

&{\footnotesize $[[r , r]]=0$}\\\hline\hline

\end{tabular}

\newpage

For three dimensional Lie super-bialgebras with the form $(\bf g
, {\tilde {\cal G}}_{\alpha\beta\gamma})$ where ${\bf g}=C^2_p,
C^3, C^4$, $C^5_{p}$ and ${\tilde {\cal G}}_{\alpha\beta\gamma}$
is one of the dual Lie superalgebras of table 4, we have table 7.
\smallskip

We see that for $p=0$ the Lie super-bialgebras $(C^2_p ,
\cal{\tilde {G}}_{\alpha\beta\gamma})$ are bi-r-matrix only for
$\beta =\gamma=0$ and $\alpha\neq 0$ (i.e., for ${\tilde {\cal
G}}_{\alpha \beta \gamma}=(A_{1,1}+2A)^0_{\epsilon,0,0}$ of table
4). Note that $(C^2_{p=0} ,(A_{1,1}+2A)^0_{\epsilon,0,0})$ are
triangular while their duals are quasi-triangular. Furthermore,
Lie superalgebras $(C^2_{p=0} ,{\tilde {\cal G}}_{\alpha, \beta,
0})$ (i.e., ${\tilde {\cal G}}_{\alpha, \beta,
0}=I_{(1,2)},\;(A_{1,1}+2A)^2_{0,1,0},\;(A_{1,1}+2A)^2_{\epsilon,1,0})$
are triangular. For $p=-1$, the Lie super-bialgebras $(C^2_p ,
{\tilde {\cal G}}_{\alpha \beta \gamma})$ with $\beta=0$ and
$\alpha, \gamma \neq 0$ (i.e., ${\tilde {\cal G}}_{\alpha \neq 0,
\beta=0, \gamma \neq 0}=(A_{1,1}+2A)^1_{\epsilon,0,\epsilon}\;,
\;(A_{1,1}+2A)^2_{\epsilon,0,-\epsilon})$ are bi-r-matrix where
$(C^2_{p=-1} ,(A_{1,1}+2A)^1_{\epsilon,0,\epsilon})$  and
$(C^2_{p=-1} ,(A_{1,1}+2A)^2_{\epsilon,0,-\epsilon})$ are
triangular while their duals are quasi-triangular. Note that for
$\beta=\gamma=0$ the Lie super-bialgebras $(C^2_{p=-1}
,(A_{1,1}+2A)^0_{\epsilon,0,0})$ are triangular $(r_2)$.
Furthermore to obtain super skew-symmetric  Poisson superbrackets
we must put $b=c$ in $r_2$.  For $p\in (-1,1]-\{0\}$ with
$\beta=0$ and $\alpha, \gamma\neq 0$ the Lie super-bialgebras
$(C^2_{p} ,(A_{1,1}+2A)^1_{\epsilon,0,\epsilon})$ and $(C^2_{p}
,(A_{1,1}+2A)^2_{\epsilon,0,-\epsilon})$ are bi-r-matrix where
they are triangular type $(r_3)$ while their duals are
quasi-triangular $({\tilde r}_2)$; on the other hand  the Lie
super-bialgebras $(C^2_{p} ,I_{(1,2)}),\;(C^2_{p}
,(A_{1,1}+2A)^0_{\epsilon,0,0}),\;(C^2_{p}
,(A_{1,1}+2A)^0_{0,0,\epsilon}),\;(C^2_{p}
,(A_{1,1}+2A)^0_{\epsilon,\epsilon,\epsilon}),\;(C^2_{p}
,(A_{1,1}+2A)^1_{\epsilon,k\neq 0,\epsilon}),\;(C^2_{p}
,(A_{1,1}+2A)^2_{0,1,0}),\;(C^2_{p}
,(A_{1,1}+2A)^2_{\epsilon,1,0}),\;(C^2_{p}
,(A_{1,1}+2A)^2_{0,1,\epsilon})$ and $(C^2_{p}
,(A_{1,1}+2A)^2_{\epsilon,k\neq 0,-\epsilon})$ are triangular.

\smallskip

Note that  $(C^3 , I_{(1,2)})$ is triangular Lie super-bialgebras,
while $(C^3,(A_{1,1}+2A)^0_{1,0,0})$ and
$(C^3,(A_{1,1}+2A)^2_{0,\epsilon,0})$ are triangular bi-r-matrix
Lie super-bialgebras with quasi-triangular duals. Furthermore, for
obtaining super skew-symmetric  Poisson superbrackets for
$(C^3,(A_{1,1}+2A)^0_{1,0,0})$ we must put  $c=-\frac{1}{2}$ in
$r$ and for $(C^3,(A_{1,1}+2A)^2_{0,\epsilon,0})$ we must put
$m=\eta=0$ in $\tilde r$.

\smallskip

Note that the Lie super-bialgebras  $(C^4 , {\tilde {\cal
G}}_{\alpha\beta\gamma})$ for $\alpha=\gamma=0$ and $\beta\neq 0$
(i.e., ${\tilde {\cal G}}_{\alpha=0, \beta, \gamma=0}=
(A_{1,1}+2A)^2_{0,\epsilon,0})$ are bi-r-matrix of triangular type
while their duals  are quasi-triangular. For $\beta=0$, $(C^4 ,
{\tilde {\cal G}}_{\alpha \beta \gamma})$  (i.e., ${\tilde {\cal
G}}_{\alpha, \beta,
\gamma}=I_{(1,2)},\;(A_{1,1}+2A)^0_{\epsilon,0,0}\;,
\;(A_{1,1}+2A)^0_{0,0,\epsilon},\;(A_{1,1}+2A)^1_{k,0,1}\;,
\;(A_{1,1}+2A)^1_{s,0,-1},\;(A_{1,1}+2A)^2_{k,0,1}\;,
\;(A_{1,1}+2A)^2_{s,0,-1})$ are triangular Lie super-bialgebras.

\smallskip

The Lie super-bialgebras  $(C^5_p , {\tilde {\cal
G}}_{\alpha,\beta=0,\gamma})$ with $p=0,\;\alpha+\gamma=0$ and
$\gamma \neq 0$ (i.e., ${\tilde {\cal G}}_{\alpha, \beta=0,
\gamma}= (A_{1,1}+2A)^2_{k,0,1},\;(A_{1,1}+2A)^2_{s,0,-1})$ are
bi-r-matrix $(r_1, {\tilde r})$ of type triangular while their
duals are quasi-triangular. Furthermore,  to obtain super
skew-symmetric Poisson  superbrackets we must put
$b=\frac{\gamma}{2}$ in $r_1$. Note that for
$p>0,\;\alpha+\gamma=0$ and $\gamma \neq 0$ the Lie
super-bialgebras $(C^5_p , (A_{1,1}+2A)^2_{k,0,1}),\;(C^5_p ,
(A_{1,1}+2A)^2_{s,0,-1})$ are bi-r-matrix $(r_2, {\tilde r})$ of
type triangular while their duals are quasi-triangular. Meanwhile
for $p>0$ and $ \alpha, \beta, \gamma = 0, 1, -1, \epsilon, k, s$
(i.e., ${\tilde {\cal G}}_{\alpha \beta \gamma}=I_{(1,2)}\;,
\;(A_{1,1}+2A)^0_{0,0,\epsilon},\;(A_{1,1}+2A)^1_{k,0,1}\;,
\;(A_{1,1}+2A)^1_{s,0,-1},\;(A_{1,1}+2A)^2_{k,0,1}\;,
\;(A_{1,1}+2A)^2_{s,0,-1})$ are triangular.
\smallskip

Finally  the Lie super-bialgebras $((A_{1,1}+2A)^0 ,
I_{(1,2)}),\;((A_{1,1}+2A)^1 , I_{(1,2)})$ and $((A_{1,1}+2A)^2 ,
I_{(1,2)})$ are triangular. Note that  for obtaining super
skew-symmetric  Poisson  superbrackets we must put $a=0,
\eta=-\zeta$ for Lie super-bialgebra $((A_{1,1}+2A)^0 ,
I_{(1,2)})$ i.e., $r=\zeta X_1 \wedge X_3 + \frac{d}{2}X_3 \wedge
X_3$ and for $((A_{1,1}+2A)^1 , I_{(1,2)})$ we must put $a=0$
(i.e., $r=0$) and for $((A_{1,1}+2A)^2 , I_{(1,2)})$ we must put
$a=b=0$ (i.e., $r=0$); where in the two later cases we have
trivial solutions.
\smallskip

Note that all these coboundary Lie super-bialgebras are
non-isomorphic. In the previous section, we mentioned the
condition (8) under which the coboundary Lie super-bialgebras are
isomorphic. Here, we consider these conditions in a more exact and
non-formal way. Using the matrix form of the isomorphism map
$\alpha: \bf g \longrightarrow {\bf g}^\prime$
\begin{equation}
\alpha(X_i) = (-1)^j\;\alpha_i^{\; \; j} {X^\prime}_j,
\end{equation}
relation (8) can be rewriten as
\begin{equation}
\Big[{{{\cal X}^\prime}_i}^{st}\Big((-1)^{l+k+kj}{\alpha}^{st}
r^{st} \alpha -(-1)^{k+lk} {r^\prime}^{st} \Big)\Big]^{st} =
-{{{\cal X}^\prime}_i}^{st} \Big[ (-1)^{n+j+jk}{\alpha}^{st} r
\alpha - (-1)^{j+jn}r^\prime\Big],
\end{equation}
where in the left hand side the indices $k$ and $j$ are row and
column of matrix $r^{st}$   respectively and $l$ corresponds to
the column of matrix $\alpha$. In the right hand side the indices
$j$ and $k$ are row and column of matrix $r$ respectively  and
$n$ denotes the column of matrix $\alpha$. If the above matrix
relations are satisfied then the two coboundary Lie
super-bialgebras $({\bf g},\tilde{\bf g})$ and $({\bf
g}^\prime,{\tilde{\bf g}}^\prime)$ are isomorphic. In this way,
one can investigate these relations for all coboundary Lie
super-bialgebras of tables and see that all of them are
non-isomorphic.

\newpage
\section {\large {\bf Calculation of super Poisson structures by Sklyanin
superbracket}}

For the triangular and quasitriangular Lie super-bialgebras one
can obtain the corresponding Poisson-Lie supergroups by means of
 Sklyanin superbracket provided by a given  super  skew-symmetric
r-matrix $r=r^{ij}X_i \otimes X_j$ \cite{{N.A},{J.z}} {\small
\begin{equation}
\{f\;,\;h\} = f \frac{{\overleftarrow{\partial}}}{\partial
x^{\mu}}\;{^\mu\hspace{-0.5mm}{X^{(L,\;r) }_i}}\; r^{ij}
{{_jX}^{(L,\;l) }}^\nu \frac{{\overrightarrow{\partial}}}
{\partial x^{\nu}}h-f \frac{{\overleftarrow{\partial}}}{\partial
x^{\mu}}\;{^\mu\hspace{-0.5mm}{X^{(R,\;r) }_i}}\; r^{ij}
{{_jX}^{(R,\;l) }}^\nu \frac{{\overrightarrow{\partial}}}
{\partial x^{\nu}}h,    \quad \forall f , h \in C^\infty(G),
\end{equation}}
where ${_jX}^{(L,\;l) }( X^{(L,\;r) }_i)$ and ${_jX}^{(R,\;l)
}(X^{(R,\;r) }_i)$ are left and right invariant sueprvector fields
with left (right) derivations on the Poisson-Lie supergroup $G$.
If $r$ is a solution of (CYBE), the following superbrackets are
also super Poisson structures on the supergroup $G$ \cite{N.A}
\begin{equation}
{\{f\;,\;h\}}^L = f \frac{{\overleftarrow{\partial}}}{\partial
x^{\mu}}\;{^\mu\hspace{-0.5mm}{X^{(L,\;r) }_i}}\; r^{ij}
{{_jX}^{(L,\;l) }}^\nu \frac{{\overrightarrow{\partial}}}
{\partial x^{\nu}}h,
\end{equation}
\vspace{-4mm}
\begin{equation}
\{f\;,\;h\}^R = f \frac{{\overleftarrow{\partial}}}{\partial
x^{\mu}}\;{^\mu\hspace{-0.5mm}{X^{(R,\;r) }_i}}\; r^{ij}
{{_jX}^{(R,\;l) }}^\nu \frac{{\overrightarrow{\partial}}}
{\partial x^{\nu}}h.
\end{equation}
For calculation of the left and right invariant supervector fields
on the supergroup $G$, it is enough to determine the left and
right invariant one forms. For $g\in G$ we have \cite{Luki}
\begin{equation}
g^{-1} dg = L^i\;{_iX}=(-1)^i {L^{(r)i}}\;_ \mu\;\overleftarrow
{dx^\mu}\;X_i=(-1)^i \overrightarrow {dx^\mu}\; {_\mu
L}^{(l)i}\;X_i,
\end{equation}
\begin{equation}
dg\;g^{-1}= R^i\;{_iX}=(-1)^i{R^{(r)i}}\;_\nu\;\overleftarrow
{dx^\nu}\;X_i=(-1)^i \overrightarrow {dx^\nu}\; {_\nu
R}^{(l)i}\;X_i,
\end{equation}
where $x^\mu$ are coordinates on the supergroup. Now using
\cite{D}
\begin{eqnarray}
<\overrightarrow{_ie}\;,\;\overrightarrow {dx^j}>= <
\frac{{\overrightarrow{\partial}}}{\partial x^i } \;,\;
\overrightarrow {dx^j}>={_i\delta}^j,\qquad <\overleftarrow
{dx^i}\;,\;\overleftarrow{e_j}>=<\overleftarrow {dx^i}\;,\;
\frac{{\overleftarrow{\partial}}}{\partial x^j }>={^i\delta}_j,
\end{eqnarray}
in the following relations
\begin{eqnarray}
<{_jX}^{(L(R),\;l)}\;,\;{L(R)}^{(l)i} >={_j\delta}^i, \qquad
\qquad <L(R)^{(r)i}\;,\; X_j^{(L(R),\;r)} >={^i\delta}_j,
\end{eqnarray}
where
\begin{eqnarray}
{_jX}^{(L(R),\;l)}:={{_jX}^{(L(R),\;l) }}^\mu\;
\frac{{\overrightarrow{\partial}}}{\partial x^\mu },\qquad \qquad
 X_j^{(L(R),\;r)}:= \frac{{\overleftarrow{\partial}}}{\partial x^\nu }\;\; {{^\nu
 X}_j^{(L(R),\;r)}},
\end{eqnarray}
we obtain the following results
\begin{equation}
{{_jX}^{(L(R),\;l) }}^\nu={{{_jL(R)}^{(l) }}}^{-1\;\nu},\qquad
\qquad {^\eta\hspace{-0.5mm}{X^{(L(R),\;r) }_j}} ={
^\eta\hspace{-0.5mm}L(R)}^{(r){-1}}_{\;j},
\end{equation}
To calculate the above matrices, we assume the following
parameterization of the two dimensional supergroup $G$:
\begin{equation}
g = e^{x X_1} e^{\psi X_2},
\end{equation}
in the same way, for three dimensional supergroups (with two
bosonic and one fermionic  generators) we assume
\begin{equation}
g = e^{x X_1}e^{y X_2} e^{\psi X_3},
\end{equation}
and for three dimensional supergroups (with two fermionic and one
bosonic  generators) we use the following parameterization
\begin{equation}
g = e^{x X_1}e^{\psi X_2} e^{\chi X_3}.
\end{equation}
The results are summarized in tables 8$-$10. Note that to
obtaining these results for dual Lie supergroups $\tilde G$ we
must use the following relations [the reason why we cannot use
the above formulas (28)-(33) for dual Lie supergroup $\tilde G$
is that for dual Lie superalgebras we use basis with upper indices
and in DeWitt notations these are different to lower indices for
Lie superalgebras.]

\vspace{12mm}

$$
{\tilde g}^{-1} d{\tilde g} ={\tilde L}_i\;{\tilde X}^i = {\tilde
L}^{(r)}_{i\; \mu}\;\overleftarrow {d{\tilde x}^\mu} \;{\tilde
X}^i=\overrightarrow {d{\tilde x}^\mu}\;{_\mu{\tilde
L}^{(l)}_i}\;{\tilde X}^i,
$$
\begin{equation}
d{\tilde g}\;{\tilde g}^{-1}= {\tilde R}_i\;{\tilde X}^i={\tilde
R}^{(r)}_{i\; \nu}\;\overleftarrow {d{\tilde x}^\nu} \;{\tilde
X}^i=\overrightarrow {d{\tilde x}^\nu}\;{_\nu{\tilde
R}^{(l)}_i}\;{\tilde X}^i,
\end{equation}
\begin{eqnarray}
<{^j{\tilde X}^{(L(R),\;l)}}\;,\;{\tilde L_i(\tilde R_i)}^{(l)}
>={^j\delta}\;_i, \qquad \qquad <{\tilde L_i(\tilde R_i)}^{(r)}\;,\; {\tilde
X}^{j(L(R),\;r)}
>={\delta_i}^{\;j},
\end{eqnarray}
\begin{eqnarray}
{^j{\tilde X}^{(L( R),\;l)}}:={{^j{\tilde X}^{(L( R),\;l)\mu}}
}\; \frac{{\overrightarrow{\partial}}}{\partial {\tilde x}^\mu },
\qquad \qquad {\tilde X}^{j(L(R),\;r)}:=
\frac{{\overleftarrow{\partial}}}{\partial{\tilde x}^\nu }\;\;
{{^\nu \tilde X}^{j(L(R),\;r)}},
\end{eqnarray}
\begin{eqnarray}
{{^j{\tilde X}^{\mu(L(R),\;l)}} }=\hspace{-1mm} ^j{{\tilde
L(\tilde R)}^{(l)^{-1}\mu}},\qquad \qquad{{^\mu
X}^{j(L(R),\;r)}}=\hspace{-1mm} ^\mu{{\tilde L (\tilde
R)}}^{(r)^{-1} j}.
\end{eqnarray}
Note that the Lie supergroup parameterizations for $\tilde G$ are
similar with $G$ (i.e., the parameters $x, y, \psi, \chi$ are
replaced with $\tilde x, \tilde y, \tilde \psi, \tilde \chi$).

\vspace{12mm}

{\bf Table 8} :~ {\small Left and right invariant supervector
fields over two dimensional Lie supergroups}.\\
\begin{tabular}{l l l l l l p{35mm} }
 \hline\hline\vspace{1mm}\smallskip

\hspace{3mm}${\bf g}$&\hspace{3mm}{\footnotesize $
B$}&\hspace{3mm}\hspace{1cm}$\tilde{\bf
g}$&\hspace{3mm}{\footnotesize
$(A_{1,1}+A)$}&\hspace{3mm}{\footnotesize $(A_{1,1}+A).i$}
\\ \hline

\vspace{2mm}
{\footnotesize $\left(\begin{array}{c} {_1X}^{(L,l)}\\
{_2X}^{(L,l)} \end{array} \right)$}&

{\footnotesize
$\left(\begin{array}{c}\frac{{\overrightarrow{\partial}}}
{\partial x}-\psi \frac{{\overrightarrow{\partial}}}{\partial \psi}\\
-\frac{{\overrightarrow{\partial}}}{\partial \psi}\end{array}
\right)$}&

\hspace{1cm}{\footnotesize $\left(\begin{array}{c} {^1{\tilde X}^{(L,\;l)}}\\
{^2{\tilde X}^{(L,\;l)}} \end{array} \right)$}&

{\footnotesize
$\left(\begin{array}{c}\frac{{\overrightarrow{\partial}}}
{\partial {\tilde x}}\\
\frac{{\tilde \psi}}{2}\frac{{\overrightarrow{\partial}}}
{\partial {\tilde x}}+\frac{{\overrightarrow{\partial}}}{\partial
{\tilde \psi}}\end{array} \right)$}&

{\footnotesize
$\left(\begin{array}{c}\frac{{\overrightarrow{\partial}}}
{\partial {\tilde x}}\\
-\frac{{\tilde \psi}}{2}\frac{{\overrightarrow{\partial}}}
{\partial {\tilde x}}+\frac{{\overrightarrow{\partial}}}{\partial
{\tilde \psi}}\end{array}
\right)$}\\
\hline
\end{tabular}

\newpage

{\bf Table 9} :~{\small Left and right invariant
supervector fields over three dimensional Lie supergroups}.\\
\begin{tabular}{l l l l l p{1mm} }
 \hline\hline \smallskip

\hspace{3mm}${\bf g}$&\hspace{3mm}{\footnotesize $
(2A_{1,1}+A)$}&\hspace{3mm}{\footnotesize
$B+A_{1,1}$}&\hspace{-0.9cm}{\footnotesize $C^1_{p}\;\;(p\in
\Re)$}&\hspace{-3.1cm}{\footnotesize $(C^1_\frac{1}{2}) $}
\\ \hline

$\left(\begin{array}{c} {_1X}^{(L,l)}\\
{_2X}^{(L,l)}\\{_3X}^{(L,l)} \end{array} \right)$ &

{\footnotesize $\left(\begin{array}{c}
\frac{{\overrightarrow{\partial}}} {\partial x}\\
\frac{{\overrightarrow{\partial}}}{\partial y}\\
-\frac{\psi}{2} \frac{{\overrightarrow{\partial}}} {\partial x}
-\frac{{\overrightarrow{\partial}}}{\partial \psi}\end{array}
\right)$}&

{\footnotesize $\left(\begin{array}{c}
\frac{{\overrightarrow{\partial}}}{\partial x}-\psi
\frac{{\overrightarrow{\partial}}}{\partial \psi}\\
\frac{{\overrightarrow{\partial}}}{\partial y}\\
-\frac{{\overrightarrow{\partial}}}{\partial \psi}
\end{array} \right)$}&

\hspace{-1cm}{\footnotesize
$\left(\begin{array}{c}\frac{{\overrightarrow{\partial}}}
{\partial x}-
y\frac{{\overrightarrow{\partial}}}{\partial y}-p \psi \frac{{\overrightarrow{\partial}}}{\partial \psi}\\
\frac{{\overrightarrow{\partial}}}{\partial y}\\
-\frac{{\overrightarrow{\partial}}}{\partial \psi}\end{array}
\right)$}&

\hspace{-3.5cm}{\footnotesize $\left(\begin{array}{c}
\frac{{\overrightarrow{\partial}}} {\partial x}-
y\frac{{\overrightarrow{\partial}}}{\partial y}- \frac{\psi}{2}
\frac{{\overrightarrow{\partial}}}{\partial
\psi}\\\frac{{\overrightarrow{\partial}}}{\partial y}\\
-\frac{\psi}{2} \frac{{\overrightarrow{\partial}}}{\partial
y}-\frac{{\overrightarrow{\partial}}}{\partial \psi}\end{array}
\right)$}\\
\end{tabular}
\vspace{5cm}

{\bf Table 10} : ~{\small Left and right invariant
supervector fields over three dimensional Lie supergroups}.\\
\begin{tabular}{l l l l l p{1mm} }
 \hline\hline \smallskip

\hspace{3mm}$\tilde {\bf g}$&\hspace{3mm}{\footnotesize $
B+A_{1,1}|.i$}&\hspace{3mm}{\footnotesize
$(2A_{1,1}+A)$}&{\footnotesize
$(2A_{1,1}+A).i$}&\hspace{-1cm}{\footnotesize
$C^1_{-p}.i\;\;(p\in\Re)$}
\\ \hline

\vspace{2mm}

$\left(\begin{array}{c} {^1{\tilde X}^{(L,\;l)}}\\
{^2{\tilde X}^{(L,\;l)}}\\{^3{\tilde X}^{(L,\;l)}} \end{array}
\right)$ &

{\footnotesize $\left(\begin{array}{c}
\frac{{\overrightarrow{\partial}}} {\partial {\tilde x}}\\
\frac{{\overrightarrow{\partial}}}{\partial {\tilde y}}-{\tilde
\psi}\frac{{\overrightarrow{\partial}}} {\partial {\tilde \psi}}\\
\frac{{\overrightarrow{\partial}}}{\partial {\tilde
\psi}}\end{array} \right)$}&

{\footnotesize $\left(\begin{array}{c}
\frac{{\overrightarrow{\partial}}} {\partial {\tilde x}}\\
\frac{{\overrightarrow{\partial}}}{\partial {\tilde y}}\\
\frac{\tilde \psi}{2} \frac{{\overrightarrow{\partial}}} {\partial
{\tilde x}} +\frac{{\overrightarrow{\partial}}}{\partial {\tilde
\psi}}\end{array} \right)$}&

{\footnotesize $\left(\begin{array}{c}
\frac{{\overrightarrow{\partial}}} {\partial {\tilde x}}\\
\frac{{\overrightarrow{\partial}}}{\partial {\tilde y}}\\
-\frac{\tilde \psi}{2} \frac{{\overrightarrow{\partial}}}
{\partial {\tilde x}}
+\frac{{\overrightarrow{\partial}}}{\partial {\tilde
\psi}}\end{array} \right)$}&

\hspace{-1cm}{\footnotesize $\left(\begin{array}{c} e^{-\tilde
y}\frac{{\overrightarrow{\partial}}}{\partial {\tilde
x}}\\\frac{{\overrightarrow{\partial}}}{\partial {\tilde y}}-p
\tilde \psi \frac{{\overrightarrow{\partial}}}{\partial \tilde
\psi}\\\frac{{\overrightarrow{\partial}}}{\partial \tilde \psi}
\end{array} \right)$}\\
\end{tabular}

\vspace{1mm}
Now using these results we can calculate the super Poisson
structures over the Lie supergroup $G$ and $\tilde G$. We can
simplify rewrite relation (25) in the following matrix form
{\footnotesize
$$
\{f\;,\;g\}=\left( \begin{array}{ccc}
                     f{\overleftarrow{X_1}}^{(L,r)} & f{\overleftarrow{X_2}}^{(L,r)} & f{\overleftarrow{X_3}}^{(L,r)}\\
                      \end{array} \right) r \left( \begin{array}{c}
                      {\overrightarrow{_1X}}^{(L,l)}g \\ {\overrightarrow{_2X}}^{(L,l)}g \\{\overrightarrow{_3X}}^{(L,l)}g\\
                     \end{array} \right)$$}
\vspace{-2mm} {\footnotesize \begin{equation}
                    \hspace{1.5cm} - \left( \begin{array}{ccc}
                      f{\overleftarrow{X_1}}^{(R,r)} & f{\overleftarrow{X_2}}^{(R,r)} & f{\overleftarrow{X_3}}^{(R,r)}\\
                      \end{array} \right) r \left( \begin{array}{c}
                      {\overrightarrow{_1X}}^{(R,l)}g \\ {\overrightarrow{_2X}}^{(R,l)}g \\{\overrightarrow{_3X}}^{(R,l)}g\\
                      \end{array} \right),
\end{equation}}
and  we can similarly rewrite (26) and (27).\footnote {Note that
for dual Lie supergroup $\tilde G$ we use the following Sklyanin
superbracket
$$
\{{\tilde f}\;,\;{\tilde h}\} = (-1)^i\;\Big({\tilde f}
\frac{{\overleftarrow{\partial}}}{\partial {\tilde x}^{\mu}}
\;{{^\mu \tilde X}^{i(L,\;r)}}\; {\tilde r}_{ij}\; {{^j{\tilde
X}^{(L,\;l)\nu}}} \frac{{\overrightarrow{\partial}}} {\partial
{\tilde x}^{\nu}}{\tilde h}-{\tilde f}
\frac{{\overleftarrow{\partial}}}{\partial {\tilde
x}^{\mu}}\;{{^\mu \tilde X}^{i(R,\;r)}}\; {\tilde r}_{ij}\;
{{^j{\tilde X}^{(R,\;l)\nu}} }\frac{{\overrightarrow{\partial}}}
{\partial {\tilde x}^{\nu}}{\tilde h}\Big),\quad \forall {\tilde
f} , {\tilde h} \in C^\infty(\tilde G)
$$}
In this manner, we calculate the fundamental Poisson
superbrackets of all triangular and quasitriangular Lie
super-bialgebras. The results are given in tables 11$-$14. Note
that for triangular Lie super-bialgebras we have calculated all
super Poisson structures (25)$-$(27)  and  listed them in
separate tables.

\vspace{3mm}

\newpage
{\bf Table 11}:$\;\;${\small Poisson superbrackets related to the
bi-r-matrix,  triangular and quasi-triangular three dimensional
\\\vspace{-1mm} {\hspace{21mm} Lie super-bialgebras  of the type
$(2,1)$}.
\smallskip}\\
\begin{tabular}{l l l l l l l l p{40mm} }
    \hline\hline

\vspace{-2mm}

{\tiny $({\bf g} , \tilde {\bf g})$} &{\tiny
$\Big(C^1_{p},C^1_{-p}.i\Big)$}& {\tiny
$\Big(C^1_{p},C^1_{-p}.i\Big)$}& {\tiny
$\Big(C^1_{p},C^1_{-p}.i\Big)$} &\hspace{-9mm}{\tiny
$\Big(C^1_{p},C^1_{-p}.i\Big)$}&\hspace{-19mm}{\tiny
$\Big((C^1_{\frac{1}{2}}) , (C^1_{\frac{1}{2}, \epsilon
})\Big)$}&\hspace{-4mm}{\tiny $ \Big(B+A_{1,1} , B+A_{1,1}|.i\Big)$} \\

\vspace{-0.5mm}

&\hspace{2mm}{\tiny $p=o$}&\hspace{2mm}{\tiny $p=1$}&
\hspace{2mm}{\tiny $p=-1$} &\hspace{-9mm}{\tiny $p\in
\Re-\{0,1,-1\}$}&& \\
\hline

{\footnotesize $\{x,y\}^L$}&{\footnotesize $1$}&{\footnotesize
$1$}
&{\footnotesize $1$}&\hspace{-8mm} {\footnotesize $1$}&\hspace{-18mm}{\footnotesize $\epsilon$}&\hspace{-2mm}{\footnotesize $1$}\\

\vspace{0.5mm}

{\footnotesize $\{x,\psi\}^L$}& {\footnotesize
$0$}&{\footnotesize $0$}&{\footnotesize $0$}
&\hspace{-7mm}{\footnotesize $0$}&\hspace{-18mm}{\footnotesize $0$}&\hspace{-2mm}{\footnotesize $0$}\\
\vspace{0.5mm}

{\footnotesize $\{y,\psi\}^L$} & {\footnotesize $0$} &
{\footnotesize $\psi-\zeta$}&
{\footnotesize $-\psi$}&\hspace{-7mm}{\footnotesize $p \psi$}&\hspace{-18mm}{\footnotesize $\epsilon \psi$}&\hspace{-2mm}{\footnotesize $\psi$}\\
\vspace{0.5mm}

{\footnotesize $\{\psi,\psi\}^L$} &{\footnotesize $a$}&
{\footnotesize $0$} & {\footnotesize $0$}& \hspace{-7mm}{\footnotesize $0$ }&\hspace{-18mm}{\footnotesize $-\epsilon$}&
\hspace{-2mm}{\footnotesize $0$}\\
\vspace{0.5mm}

{\footnotesize  $\{x,y\}^R$} & {\footnotesize $e^{-x}$}
&{\footnotesize $e^{-x}$}
&{\footnotesize $e^{-x}$}& \hspace{-7mm}{\footnotesize $e^{-x}$}&\hspace{-18mm}{\footnotesize $\epsilon e^{-x}$}&\hspace{-2mm}{\footnotesize $1$}\\
\vspace{0.5mm}

{\footnotesize  $\{x,\psi\}^R$} & {\footnotesize $0$}&
{\footnotesize $0$} &{\footnotesize $0$}
&\hspace{-7mm}{\footnotesize $0$}&\hspace{-18mm}{\footnotesize $0$}&\hspace{-2mm}{\footnotesize $0$}\\
\hline\hline

\end{tabular}
\newpage
{\bf Table 12}:~~{\small Poisson superbrackets related to the
triangular and quasi-triangular three dimensional
\\\vspace{-1mm} {\hspace{21mm} Lie
super-bialgebras of the type $(2,1)$}.
\smallskip}\\
\begin{tabular}{l l l l l  p{-2mm} }
\hline\hline
\vspace{1mm}

{\footnotesize $({\bf g} , \tilde {\bf g})$}&{\footnotesize
$\Big((2A_{1,1}+A) , I_{(2,1)}\Big)$} &
{\footnotesize$\Big(C^1_{p=0} , I_{(2,1)}\Big)$}&
{\footnotesize$\Big(C^1_{p=1} , I_{(2,1)}\Big)$}
&\hspace{-1cm}{\footnotesize
$\Big(C^1_{p=1} , (2A_{1,1}+A)\Big)$} \\
\hline

\vspace{1mm}

{\footnotesize $\{x,y\}^L$}& {\footnotesize $b$}&{\footnotesize
$0$}
&{\footnotesize $0$}&\hspace{-1.15cm} {\footnotesize $0$}\\

\vspace{1mm}

{\footnotesize $\{x,\psi\}^L$}& {\footnotesize
$0$}&{\footnotesize $0$}
&{\footnotesize $0$}&\hspace{-1cm}{\footnotesize $0$}\\

 \vspace{1mm}

{\footnotesize $\{y,\psi\}^L$} &{\footnotesize $0$} &
{\footnotesize $0$}&
{\footnotesize $-\zeta$}&\hspace{-1cm}{\footnotesize $-\zeta$}\\

\vspace{1mm}

{\footnotesize $\{\psi,\psi\}^L$} &{\footnotesize $0$}
 & {\footnotesize $a$}& {\footnotesize $0$ }&\hspace{-1cm}{\footnotesize $-\frac{1}{2}$}\\

\vspace{1mm} {\footnotesize  $\{x,y\}^R$} & {\footnotesize $b$}&
{\footnotesize $0$}
&{\footnotesize $0$}&\hspace{-1cm}{\footnotesize $0$}\\

\vspace{1mm} {\footnotesize  $\{x,\psi\}^R$} & {\footnotesize
$0$}&{\footnotesize $0$}
&{\footnotesize $0$}&\hspace{-1cm}{\footnotesize $0$}\\

\vspace{1mm}

{\footnotesize  $\{y,\psi\}^R$} & {\footnotesize $0$}&
{\footnotesize $0$}
&{\footnotesize $-\zeta e^{-2x}$}&\hspace{-1cm}{\footnotesize $-\zeta e^{-2x}$}\\
\hline\hline
\end{tabular}

\bigskip
\bigskip
{\bf Table 13}:~ {\small Poisson superbrackets related to the
triangular and quasi-
\\\vspace{-1mm} {\hspace{15mm} triangular
two dimensional Lie super-bialgebras of the type $(1,1)$}.
\smallskip}\\
\begin{tabular}{l l l l p{100mm} }
    \hline\hline\vspace{1mm}

\smallskip
{\footnotesize $({\bf g} , \tilde {\bf
g})$}&\hspace{2cm}{\footnotesize $\Big(B , (A_{1,1}+A)\Big)$}&\hspace{2cm}{\footnotesize$\Big(B , (A_{1,1}+A).i\Big)$} \\
\hline

\vspace{1mm}

{\footnotesize $\{x,\psi\}^L$}&\hspace{2cm} {\footnotesize $0$}&\hspace{2cm}{\footnotesize $0$}\\

 \vspace{1mm}

{\footnotesize $\{\psi,\psi\}^L$} &\hspace{2cm} {\footnotesize
$-\frac{1}{2}$} &\hspace{2cm}
{\footnotesize $\frac{1}{2}$}\\

\vspace{1mm}
{\footnotesize  $\{x,\psi\}^R$} &\hspace{2cm} {\footnotesize $0$} &\hspace{2cm}{\footnotesize $0$}\\

\vspace{1mm}

{\footnotesize  $\{\psi,\psi\}^R$} & \hspace{2cm}
{\footnotesize  $-\frac{1}{2} e^{-2x}$}&\hspace{2cm}{\footnotesize $\frac{1}{2} e^{-2x}$}  \\

\vspace{1mm}
{\footnotesize  $\{x,\psi\}$} &\hspace{2cm} {\footnotesize  $0$} &\hspace{2cm}{\footnotesize $0$}\\

\vspace{1mm}

{\scriptsize $\{\psi,\psi\}$} &\hspace{2cm} {\footnotesize
$-\frac{1}{2}(1-e^{-2x})$} & \hspace{2cm}{\footnotesize
$\frac{1}{2}(1-e^{-2x})$}\\

\vspace{1mm}

{\footnotesize  $\{\tilde x,\tilde \psi\}$} & \hspace{2cm}
{\footnotesize  $-\tilde \psi$} &\hspace{2cm}{\footnotesize $-\tilde \psi$}\\

\vspace{1mm}

{\scriptsize $\{\tilde \psi,\tilde \psi\}$} &\hspace{2cm}
{\footnotesize $0$} & \hspace{2cm}{\footnotesize $0$}
\smallskip \\
\hline\hline
\end{tabular}

\vspace{12mm}

{\bf Table 14}:~{\small Poisson superbrackets related  to the
bi-r-matrix, triangular  three dimensional\\\vspace{0.5mm}
\hspace{23mm} Lie super-bialgebras of the type $(1,2)$.\footnote {
Note that the Poisson
superbrackets exist if the conditions in the comments are satisfied.}}\\
\begin{tabular}{l l l l l l l l p{20mm} }
    \hline\hline
\hspace{1mm}

{\footnotesize $({ \bf g} , \tilde {\bf g})$} & {\footnotesize
$(C^2_{p=0} ,{\cal{\tilde {G}}}_{\alpha,\beta,0 })$}&\hspace{-1mm}
{\scriptsize Comments} &{\footnotesize $(C^2_{p=-1} ,{\cal{\tilde
{G}}}_{\alpha,0,\gamma})$} &\hspace{-1mm} {\scriptsize
Comments}&{\footnotesize ${(C^2_{p} ,\cal{\tilde
{G}}_{\alpha\beta\gamma})}_{|_{p\in (-1,1]-\{0\}}}$}&\hspace{-2mm}
{\scriptsize Comments}\smallskip\\\hline

\vspace{1mm}

{\footnotesize $\{x,\psi\}^L$}& {\footnotesize
$0$}&&{\footnotesize $0$}&&{\footnotesize $0$}&\\
\vspace{1mm}

{\footnotesize $\{x,\chi\}^L$} & {\footnotesize
$0$}&&{\footnotesize $0$}&& {\footnotesize $0$}&\\
\vspace{1mm}

{\footnotesize $\{\psi,\chi\}^L$}&{\footnotesize
$-\beta$}&& {\footnotesize $b$}& &{\footnotesize $-\frac{\beta}{p+1}$}&\\
\vspace{1mm}

{\footnotesize $\{\psi,\psi\}^L$}&{\footnotesize
$-\frac{\alpha}{2}$}& & {\footnotesize $-\frac{\alpha}{2}$}&&
{\footnotesize $-\frac{\alpha}{2}$}&\\
\vspace{1mm}

{\footnotesize $\{\chi,\chi\}^L$}&{\footnotesize
$b$}&&{\footnotesize $\frac{\gamma}{2}$}&& {\footnotesize $-\frac{\gamma}{2p}$}&\\
\vspace{1mm}

{\footnotesize $\{x,\psi\}^R$}& {\footnotesize
$0$}&&{\footnotesize $0$}& &{\footnotesize $0$}&\\
\vspace{1mm}

{\footnotesize $\{x,\chi\}^R$} & {\footnotesize
$0$}&& {\footnotesize $0$}& &{\footnotesize $0$}&\\
\hline
\end{tabular}
\smallskip
\goodbreak
\vspace{5mm}
{ {\small {\bf  Table 14}-Continued. }}\smallskip\\
\begin{tabular}{l l l l lp{20mm} }
    \hline

{\footnotesize $({ \bf g} , \tilde {\bf g})$} & {\footnotesize
$(C^3 ,I_{(1,2)})$} & {\footnotesize $(C^3
,(A_{1,1}+2A)^0_{1,0,0})$}&{\footnotesize $(C^4 ,\cal{\tilde
{G}}_{\alpha\beta\gamma})$}&\hspace{-5mm} {\scriptsize
Comments}\\\hline

\vspace{1mm}

{\footnotesize $\{x,\psi\}^L$}& {\footnotesize
$-\zeta$}&{\footnotesize $0$}&{\footnotesize $0$}\\
\vspace{1mm}

{\footnotesize $\{x,\chi\}^L$} & {\footnotesize
$0$}&{\footnotesize $0$}
 & {\footnotesize $0$}&\\
\vspace{1mm}

{\footnotesize $\{\psi,\chi\}^L$}&{\footnotesize
$0$}&{\footnotesize $-\frac{1}{2}$} &{\footnotesize $\frac{\gamma-2\beta}{4}$}&\\
\vspace{1mm}

{\footnotesize $\{\psi,\psi\}^L$}&{\footnotesize
$c$}&{\footnotesize $b$}&
{\footnotesize $\frac{2(\beta-\alpha)-\gamma}{4}$}&\\
\vspace{1mm}

{\footnotesize $\{\chi,\chi\}^L$}&{\footnotesize
$0$}&{\footnotesize $0$}  &{\footnotesize $-\frac{\gamma}{2}$}\\
\vspace{1mm}

{\footnotesize $\{x,\psi\}^R$}& {\footnotesize
$-\zeta$}&{\footnotesize
$0$}&{\footnotesize $0$}\\
\vspace{1mm}

{\footnotesize $\{x,\chi\}^R$} & {\footnotesize
$0$}&{\footnotesize $0$}
&{\footnotesize $0$}\\
\hline\hline
\end{tabular}
\smallskip
\goodbreak
\vspace{2cm}
{ {\small  {\bf Table 14}-Continued. }}\smallskip\\
\begin{tabular}{l  l l l lp{20mm} }
    \hline
\vspace{1mm}

{\footnotesize $({ \bf g} , \tilde {\bf g})$} & {\footnotesize
$(C^5_{p=0} ,{\cal{\tilde {G}}}_{-\gamma,0,\gamma})$}
&\hspace{-3mm} {\scriptsize Comments}& {\footnotesize $(C^5_{p>0}
,{\cal{\tilde {G}}}_{\alpha,0,\gamma})$}&\hspace{-5mm}
{\scriptsize Comments}\\\hline

\vspace{1mm}

{\footnotesize $\{x,\psi\}^L$}& {\footnotesize
$0$}&&{\footnotesize $0$}&\\
\vspace{1mm}

{\footnotesize $\{x,\chi\}^L$} & {\footnotesize
$0$}&&{\footnotesize $0$}
 & \\
\vspace{1mm}

{\footnotesize $\{\psi,\chi\}^L$}&{\footnotesize
$\frac{\gamma}{2}$}&&{\footnotesize $\frac{\gamma-\alpha}{4p(1+p^2)}$} &\\
\vspace{1mm}

{\footnotesize $\{\psi,\psi\}^L$}&{\footnotesize
$a$}&&{\footnotesize $-\frac{2\alpha p^2+\alpha+\gamma}{4p(1+p^2)}$}&\\
\vspace{1mm}

{\footnotesize $\{\chi,\chi\}^L$}&{\footnotesize
$a$}&&{\footnotesize $-\frac{\alpha+\gamma+2\gamma p^2}{4p(1+p^2)}$}&\\
\vspace{1mm}

{\footnotesize $\{x,\psi\}^R$}& {\footnotesize
$0$}&&{\footnotesize
$0$}&\\
\vspace{1mm}

{\footnotesize $\{x,\chi\}^R$} & {\footnotesize
$0$}&&{\footnotesize $0$}
&\\
\hline\hline
\end{tabular}

\smallskip
\goodbreak
\vspace{8mm}

{ {\small {\bf Table 14}-Continued. }}\smallskip\\
\begin{tabular}{l l| l l}
\hline
{\footnotesize $({ \bf g} , \tilde {\bf g})$} &
{\footnotesize $(C^3 ,(A_{1,1}+2A)^2_{0, \epsilon,
0})$}&{\footnotesize $({ \bf g} , \tilde {\bf g})$} &
{\footnotesize $((A_{1,1}+2A)^0 ,I_{(1,2)})$}\smallskip\\\hline

{\footnotesize $\{x,\psi\}$}& {\footnotesize $0$}&{\footnotesize
$\{x,\psi\}^L$}& {\footnotesize
$0$}\\

{\footnotesize $\{x,\chi\}$}& {\footnotesize $0$}&{\footnotesize
$\{x,\chi\}^L$} & {\footnotesize
$-\zeta$}\\

{\footnotesize $\{\psi,\chi\}$}&{\footnotesize $-\epsilon
x$}&{\footnotesize $\{\psi,\chi\}^L$}&{\footnotesize
$0$}\\

{\footnotesize $\{\psi,\psi\}$}&{\footnotesize $\epsilon
x^2$}&{\footnotesize $\{\psi,\psi\}^L$}&{\footnotesize
$0$}\\

{\footnotesize $\{\chi,\chi\}$}&{\footnotesize
$0$}&{\footnotesize $\{\chi,\chi\}^L$}&{\footnotesize
$d$}\\

{\footnotesize $\{\tilde x,\tilde \psi\}^L$}& {\footnotesize
$0$}&{\footnotesize $\{x,\psi\}^R$}& {\footnotesize $0$}\\

{\footnotesize $\{\tilde x,\tilde \chi\}^L$} & {\footnotesize
$0$}&{\footnotesize $\{x,\chi\}^R$} & {\footnotesize
$-\zeta$}\\
\hline\hline
\end{tabular}
\section {\large {\bf Conclusion }}
Having  determined the types (triangular,  quasi-triangular or
factorizable) of two and three dimensional Lie super-bialgebras
and obtained their r-matrices and super Poisson structures we are
now in a position to perform the quantization of these Lie
super-bialgebras. Furthermore,  one can now investigate
integrability under super Poisson-Lie T duality by studying the
super Poisson-Lie T dual sigma models \cite{ER} over  bi-r- matrix
super-bialgebras.

\bigskip
\bigskip

{\bf Acknowledgments}. The authors   thank S. Moghadassi for
carefully reading the manuscript and for the useful comments.

This work has supported by research vice chancellor of Azarbaijan
Shahid Madani University.


\end{document}